\documentclass{pasa}

\usepackage{graphicx}
\usepackage{multirow}
\usepackage{caption, subcaption}

\title[Source counts and confusion in GLEAM]{Source counts and confusion at 72--231~MHz in the MWA GLEAM survey}

\author[Franzen et al.]{T.~M.~O.~Franzen$^{1,2,3,}$\thanks{Email: franzen@astron.nl} ,
T.~Vernstrom$^{4}$,
C.~A.~Jackson$^{1,5,3}$,
N.~Hurley-Walker$^{1}$,
R.~D.~Ekers$^{1}$,
G.~Heald$^{2}$,
N.~Seymour$^{1}$,
and S.~V.~White$^{1}$
\affil{$^1$International Centre for Radio Astronomy Research, Curtin University, Bentley, WA 6102, Australia}
\affil{$^2$CSIRO Astronomy and Space Science, PO Box 1130, Bentley WA 6102, Australia}
\affil{$^3$ASTRON, Netherlands Institute for Radio Astronomy, Oude Hoogeveensedijk 4, 7991 PD, Dwingeloo, The Netherlands}
\affil{$^4$Dunlap Institute for Astronomy and Astrophysics, University of Toronto, ON, M5S 3H4, Canada}
\affil{$^5$ARC Centre of Excellence for All-sky Astrophysics (CAASTRO)}
}

\jid{PASA}
\doi{10.1017/pas.\the\year.xxx}
\jyear{\the\year}

\usepackage{aas_macros}
\usepackage{hyperref} 
\hypersetup{colorlinks,citecolor=blue,linkcolor=blue,urlcolor=blue}

\hypersetup{draft}
\begin{document}

\begin{frontmatter}
\maketitle

\begin{abstract}
The GaLactic and Extragalactic All-sky MWA survey (GLEAM) is a radio continuum survey at 72--231~MHz of the whole sky south of declination~$+ 30^{\circ}$, carried out with the Murchison Widefield Array (MWA). In this paper, we derive source counts from the GLEAM data at 200, 154, 118 and 88~MHz, to a flux density limit of 50, 80, 120 and 290~mJy respectively, correcting for ionospheric smearing, incompleteness and source blending. These counts are more accurate than other counts in the literature at similar frequencies as a result of the large area of sky covered and this survey's sensitivity to extended emission missed by other surveys. At $S_{154~\mathrm{MHz}} > 0.5$~Jy, there is no evidence of flattening in the average spectral index ($\alpha \approx -0.8$ where $S \propto \nu^{\alpha}$) towards the lower frequencies. We demonstrate that the SKA Design Study (SKADS) model by \citeauthor{wilman2008} significantly underpredicts the observed 154~MHz GLEAM counts, particularly at the bright end. Using deeper LOFAR counts and the SKADS model, we find that sidelobe confusion dominates the thermal noise and classical confusion at $\nu \gtrsim 100$~MHz due to both the limited CLEANing depth and undeconvolved sources outside the field-of-view. We show that we can approach the theoretical noise limit using a more efficient and automated CLEAN algorithm.
\end{abstract}

%We suggest that this is due to the limited volume covered by the model, biasing against rare, bright sources. 

\end{frontmatter}

%%%%%%%%%%%%%%%%%%%%%%%%%%%%%%%%%%%%%%%%%%%%%%%%%%%%%%%%%%%%%%%%%%%
\section{Introduction }\label{Introduction}

Differential radio source counts are important because they constrain the nature and evolution of extragalactic sources, and unlike luminosity functions, do not require redshifts. They have to date been best studied at 1.4~GHz. At the highest flux densities ($S \gtrsim 10$~Jy), the 1.4-GHz Euclidean normalised differential counts, $\frac{dN}{dS} S^{2.5}$, show a flattened region, as expected in a static, non-evolving (`Euclidean') Universe. Below $\sim 10$~Jy, the counts rise with decreasing flux density followed by a plateau and then a steep fall. This bulge is recognised \citep{longair1966} as an indicator of cosmic evolution, in which radio-luminous sources undergo greater evolution in comoving space density than their less-luminous counterparts. \cite{condon1984} and \cite{windhorst1985} found that the source count slope flattens around 1~mJy, suggesting a new population of radio sources at low flux densities. This new population is now widely thought to consist predominantly of star-forming galaxies with an admixture of radio-quiet AGN \citep[e.g.][]{jackson1999,massardi2010, dezotti2010}.

Our knowledge of the low-frequency sky ($\nu \lesssim 200$~MHz) is poor compared with that at 1.4~GHz, and consequently information about the low-frequency counts is more limited. Low-frequency surveys are particularly sensitive to sources with steep synchrotron spectra. They are not biased by relativistic beaming effects and favour older emission originating from the extended lobes of radio galaxies rather than emission from the core \citep{wall1994}. They therefore give a complementary view to $\sim$GHz surveys.

As well as contributing to our understanding of extragalactic source populations, low frequency counts are useful for the interpretation of Epoch of Reionisation (EoR) data, in which foreground radio sources are a critical contaminant. A number of methods to model and subtract the foreground contamination from EoR data have been explored \citep[see e.g.][]{morales2004,chapman2012,trott2012,carroll2016}. Higher resolution radio data at a similar frequency to the EoR observations can be used to directly subtract extragalactic radio sources from the EoR data while extrapolation of the known source counts can be used to model and statistically suppress sources to fainter flux densities.

Survey observations over the past few years with instruments such as the Giant Metrewave Radio Telescope \citep[GMRT;][]{swarup1991}, the Low Frequency Array \citep[LOFAR;][]{van_haarlem2013} and the Murchison Widefield Array \citep[MWA;][]{tingay2013} have provided a wealth of new information about the low-frequency sky. Recent all-sky low frequency surveys include the VLA Low-frequency Sky Survey Redux at 74~MHz \citep[VLSSr;][]{lane2014}, the Multifrequency Snapshot Sky Survey at 120--180~MHz \citep[MSSS;][]{heald2015}, the Tata Institute for Fundamental Research GMRT Sky Survey at 150~MHz \citep[TGSS;][]{intema2016} and the Galactic and Extragalactic All-sky MWA survey at 72--231~MHz \citep[GLEAM;][]{wayth2015}. Among these surveys, GLEAM has the widest fractional bandwidth and highest surface brightness sensitivity. The survey covers the entire sky south of Dec~$+ 30^{\circ}$ at an angular resolution of $\approx 2.5$~arcmin at 200~MHz and is complete to $S_{200~\mathrm{MHz}} = 50$~mJy in the deepest regions.

Much deeper and higher resolution surveys at 150~MHz covering a few tens of square degrees exist using LOFAR \citep{hardcastle2016,mahony2016,williams2016}. The deepest of these by \cite{williams2016} reaches an rms sensitivity of $\approx 120~\mu$Jy/beam. These surveys have detected a flattening in the counts below $\approx 10$~mJy which is thought to be associated with the rise of the low flux density star-forming galaxies and radio-quiet AGN, as seen at e.g. 1.4~GHz below $\approx 1$~mJy. The ongoing LOFAR Two-metre Sky Survey \citep[LoTSS;][]{shimwell2017} at 120--168~MHz will eventually cover the entire northern sky to an rms sensitivity of $\approx 100~\mu$Jy/beam.

The Square Kilometre Array Design Study (SKADS) Semi-Empirical Extragalactic Simulated Sky by \cite{wilman2008} is in wide use to facilitate predictions for the SKA sky and optimise its design and observing programmes. These models are also a valuable tool in the interpretation of existing radio surveys. The latest low frequency counts provide an opportunity to compare the model predictions and identify any deficiencies.

The confusion noise in low-frequency interferometric images is dependent on the source counts. Classical confusion occurs when the source density is so high that sources cannot be clearly resolved by the array; the image fluctuations are due to the sum of all sources in the main lobe of the synthesised beam. Sidelobe confusion introduces additional noise into an image due to the combined sidelobes of undeconvolved sources. Other basic sources of error in radio interferometric images include the system noise and calibration artefacts. It is important to analyse the relative contribution of these noise terms to assess whether enhancements in the data processing have the potential to further reduce the noise. This is also essential for statistically interpreting survey data below the source detection threshold.

\cite{franzen2016} derive the 154~MHz source counts using MWA pointed observations of an EoR field covering $570~\mathrm{deg}^{2}$, centred at J2000 $\alpha = 03^{\mathrm{h}}30^{\mathrm{m}}$, $\delta = -28^{\circ}00'$. The image has an angular resolution of 2.3~arcmin and the rms noise in the centre of the image is 4--5~mJy/beam. Using deeper GMRT source counts down to $S_{\mathrm{153~MHz}} = 6$~mJy, they estimate the classical confusion noise to be $\approx 1.7$~mJy/beam from a $P(D)$ analysis \citep{scheuer1957}. They argue that the image is limited by sidelobe confusion but they do not investigate the underlying causes of the sidelobe confusion.

In this paper, we derive the source counts to higher precision using the GLEAM survey, covering $24,831~\mathrm{deg}^{2}$, at 200, 154, 118 and 88~MHz, allowing tight constraints on bright radio source population models. We analyse any change in the shape of the source counts with frequency and compare them with the SKADS model. We use the LOFAR counts by \cite{williams2016} together with the SKADS model to derive the classical confusion noise across the entire GLEAM frequency range. We quantify the excess background noise in GLEAM and demonstrate that it is primarily caused by sidelobe confusion. We identify which aspects of the data processing contribute to sidelobe confusion and show how the sidelobe confusion can be improved. Finally, we discuss confusion limits for future MWA Phase 2 observations with the angular resolution improved by a factor of two.

%%%%%%%%%%%%%%%%%%%%%%%%%%%%%%%%%%%%%%%%%%%%%%%%%%%%%%%%%%%%%%%%%%%
\section{GLEAM observing, imaging, and source finding}\label{GLEAM observing, imaging, and source finding}

We refer the reader to \cite{wayth2015} and \cite{hurleywalker2017} for details of the survey strategy and data reduction methods for the GLEAM year 1 extragalactic catalogue respectively. In this section, we highlight the points salient to this paper.

The GLEAM survey was conducted using Phase 1 of the MWA, which consisted of 128 16-crossed-pair-dipole tiles, distributed over an area $\approx 3$~km in diameter. The whole sky south of Dec~$+30^\circ$ was surveyed using meridian drift scan observations. The sky was divided into seven declination strips and one declination strip was covered in a given night. The observing was broken into a series of 2~min scans in five frequency bands (72--103, 103--134, 139--170, 170--200 and 200--231~MHz), cycling through the five frequency bands in 10~min.

Each 2~min snapshot observation was imaged separately using \textsc{wsclean} \citep{offringa2014}, a \textit{w}-stacking deconvolution algorithm which appropriately handles the \textit{w} term for widefield imaging. For imaging purposes, the 30.72~MHz bandwidth was split into four 7.68~MHz sub-bands. The final image products consist of 20 Stokes $I$ 7.68~MHz sub-band mosaics spanning 72--231~MHz as well as four deep wide-band mosaics covering 170--231, 139--170, 103--134 and 72--103~MHz, formed by combining the 7.68~MHz sub-band mosaics.

The source finder \textsc{aegean} \citep{hancock2012,hancock2018} was run on the 170--231~MHz image to create a blind source catalogue centred at 200~MHz. The catalogue was filtered to exclude areas within $10^{\circ}$ of the Galactic plane and other areas affected by poor ionospheric conditions or containing bright, extended sources such as Centaurus A (see Table~\ref{tab:excluded_regions} for details). The filtered catalogue covers an area of $24,831~\mathrm{deg}^{2}$, hereafter referred to as region A, and contains 307,455 components above $5\sigma$, where $\sigma$ is the rms noise. It is estimated to be 90 per cent complete at $S_{200~\mathrm{MHz}} = 170$~mJy. In order to provide spectral information across the full frequency range, the priorised fitting mode of \textsc{aegean} was used to perform flux density estimates across the 20 7.68-MHz sub-bands. The catalogue provides both peak and integrated flux densities. The peak flux densities were corrected for ionospheric smearing as outlined below. The three lowest frequency wide-band images were not used to provide measurements for the catalogue.

The GLEAM flux densities are tied to the flux density scale of \cite{baars1977}. Overall, the GLEAM catalogue is consistent with \citeauthor{baars1977} to within 8 per cent for 90 per cent of the survey area, where the difference is primarily caused by uncertainty in the MWA primary beam model.

%%%%%%%%%%%%%%%%%%%%%%%%%%%%%%%%%%%%%%%%%%%%%%%%%%%%%%%%%%%%%%%%%%%
\subsection{Correcting peak flux densities for ionospheric smearing}\label{Effect of ionospheric smearing on the point spread function}

Ionospheric perturbations cause sources to be smeared out in the final, mosaicked images. The magnitude of the effect is proportional to $\nu^{-2}$, where $\nu$ is the frequency. Consequently, at any map position, the actual point spread function (PSF) is larger than the restoring beam by a certain amount, depending on the degree of ionospheric smearing. \cite{hurleywalker2017} used sources known to be unresolved in higher resolution radio surveys to sample the shape of the PSF across each of the mosaics. Maps of the variation of $a_{\mathrm{psf}}$, $b_{\mathrm{psf}}$ and $pa_{\mathrm{psf}}$ were produced, where $a_{\mathrm{psf}}$, $b_{\mathrm{psf}}$ and $pa_{\mathrm{psf}}$ are the major and minor axes and position angle of the PSF respectively. 

The increase in area of the PSF resulting from ionospheric smearing is given by 
\begin{equation}
R = \frac{a_{\mathrm{PSF}} b_{\mathrm{PSF}}}{a_{\mathrm{rst}} b_{\mathrm{rst}} } \, ,
\end{equation}
where $a_{\mathrm{rst}}$ and $b_{\mathrm{rst}}$ are the major and minor axes of the restoring beam respectively. Sources detected in the 170--231~MHz image have a mean value of $R$ of 1.14, with a standard deviation of 0.04, and in regions worst affected by ionospheric smearing, $R$ reaches 1.44. Ionospheric smearing not only increases the source area by a factor of $R$ but also reduces the peak flux density by the same amount, while integrated flux densities are preserved. In order to restore the peak flux density of the sources, the images were multiplied by $R$. In the catalogue, integrated flux densities were normalised with respect to the position-dependent PSF to ensure that, for bright point sources, peak and integrated flux densities agree.

\begin{table*}
\centering
\caption{Summary of sky regions excised from the GLEAM survey used in the analyses of this paper. The top row indicates the total surveyed area in GLEAM. The GLEAM catalogue area covers 24,831~$\mathrm{deg}^{2}$ and consists of the total surveyed area excluding the regions listed in the middle rows. The peeled sources are Hydra A, Pictor A, Hercules A, Virgo A, Crab, Cygnus A and Cassiopeia A; their positions are listed in \citet{hurleywalker2017}.}
\begin{tabular}{ccc}
\hline
Description & Region & Area ($\mathrm{deg}^2$)\\
\hline
Total surveyed area & $\mathrm{Dec}<+30^{\circ}$ & 30,940 \\
\hline
Galactic plane & Absolute Galactic latitude $<10^{\circ}$ & 4,776 \\
Ionospherically distorted & $0^{\circ}<\mathrm{Dec}<+30^{\circ}$ \& $22^\mathrm{h}<\mathrm{RA}<0^\mathrm{h}$ & 859 \\
Centaurus~A & $13^\mathrm{h}25^\mathrm{m}28^\mathrm{s} -43^{\circ}01^{\prime}09^{\prime\prime}$, $r=9^{\circ}$ & 254 \\
Sidelobe reflection of Cen~A & $13^\mathrm{h}07^\mathrm{m}<\mathrm{RA}<13^\mathrm{h}53^\mathrm{m}$ \& $20^{\circ}<\mathrm{Dec}<+30^{\circ}$ & 104 \\
Large Magellanic Cloud & $05^\mathrm{h}23^\mathrm{m}35^\mathrm{s} -69^{\circ}45^{\prime}22^{\prime\prime}$, $r=5.5^{\circ}$ & 95 \\
Small Magellanic Cloud & $00^\mathrm{h}52^\mathrm{m}38^\mathrm{s} -72^{\circ}48^{\prime}01^{\prime\prime}$, $r=2.5^{\circ}$ & 20 \\
Peeled sources & Radius of 10~arcmin & $< 1$ \\
\hline
GLEAM catalogue area (region A) & & 24,831 \\
\hline
\end{tabular}
\label{tab:excluded_regions}
\end{table*}

%\begin{table*}
%\centering
%\caption{Source finding statistics in region A, covering 24,831~$\mathrm{deg}^{2}$. For the resolution, we quote the mean and standard deviation of the PSF major axis.}
%\label{tab:source_stats}
%\begin{tabular}{@{} c c c c c c c c} 
%\hline
%\multicolumn{1}{c}{Frequency}
%&\multicolumn{1}{c}{5 $\sigma$ detection}
%&\multicolumn{1}{c}{Number}
%&\multicolumn{1}{c}{Percentage}
%&\multicolumn{1}{c}{PSF major}
%&\multicolumn{1}{c}{PSF minor}
%&\multicolumn{1}{c}{Source}
%&\multicolumn{1}{c}{Number of}\\
%(MHz) & threshold (mJy/bm) & of sources & \multicolumn{1}{c}{extended} & axis (arcsec) & axis (arcsec) & density ($\mathrm{deg}^{-2}$) & beams/source \\
%\hline
% 200 & 2 & 307,455 & 7.3 & $ 144 \pm 16 $ & $132 \pm 5$ & 12.4 & 49 \\
% 154 & 2 & 254,072 & 7.3 & $ 176 \pm 24 $ & $159 \pm 6$ & 10.2 & 40 \\
% 118 & 2 & 195,821 & 6.3 & $ 229 \pm 29 $ & $209 \pm 8$ & 7.9 & 30 \\
% 88  & 2 & 131,250 & 6.0 & $ 313 \pm 36 $ & $287 \pm 12$ & 5.3 & 24 \\
%\hline
%\end{tabular}
%\end{table*}

\begin{table*}
\centering
\caption{Source finding statistics in region A, covering 24,831~$\mathrm{deg}^{2}$. For the 5$\sigma$ detection threshold, and PSF major and minor axes, we quote the mean and standard deviation. Sources are classified as extended as described in Section~\ref{Classifying sources as point-like or extended.}}
\label{tab:source_stats}
\begin{tabular}{@{} c c c c c} 
\hline
Property & $\nu = 200$~MHz & $\nu = 154$~MHz & $\nu = 118$~MHz & $\nu = 88$~MHz \\
\hline
5$\sigma$ detection threshold (mJy/bm) & $56 \pm 37$ & $84 \pm 45$ & $137 \pm 68$ & $265 \pm 112$ \\
Number of sources & 307,455 & 254,072 & 195,821 & 131,250 \\
Percentage extended & 7.3 & 7.3 & 6.3 & 6.0 \\
PSF major axis (arcsec) & $144 \pm 16$ & $176 \pm 24$ & $229 \pm 29$ & $313 \pm 36$ \\
PSF minor axis (arcsec) & $132 \pm 5$ & $159 \pm 6$ & $209 \pm 8$ & $287 \pm 12$ \\
Source density ($\mathrm{deg}^{-2}$) & 12.4 & 10.2 & 7.9 & 5.3 \\
Number of beams/source & 49 & 40 & 30 & 24\\
\hline
\end{tabular}
\end{table*}

%%%%%%%%%%%%%%%%%%%%%%%%%%%%%%%%%%%%%%%%%%%%%%%%%%%%%%%%%%%%%%%%%%%
\section{Source finding at 154, 118 and 88~MH\lowercase{z}} \label{Source finding}

Since a statistically complete sample is required to measure the counts at any frequency, we cannot use the sub-band measurements quoted in the GLEAM catalogue, obtained from the priorised fitting, to measure the counts. In order to derive the counts at frequencies below 200~MHz, we use the wide-band images covering 139--170, 103--134 and 72--103~MHz, centred at 154, 118 and 88~MHz respectively.

We create a blind source catalogue at each of these frequencies following a similar procedure to that employed by \cite{hurleywalker2017}. We first use \textsc{bane} \citep{hancock2018} to remove the background structure and estimate the rms noise across the image. The `box' parameter defining the angular scale on which the rms and background are evaluated is set to 20 times the synthesised beam size. We then run the source finder \textsc{aegean} using a 5$\sigma$ detection threshold. The integrated flux densities are normalised using the PSF map at the relevant frequency. Sources lying within areas flagged from the GLEAM catalogue (see Table~\ref{tab:excluded_regions}) are excluded. The number of sources detected at each frequency and other source finding statistics are given in Table~\ref{tab:source_stats}.

The mosaics used to create the source catalogues have a relatively large fractional bandwidth; the 88~MHz mosaic has the largest fractional bandwidth of $\approx 0.35$. For any source with a non-zero spectral index, there is a discrepancy between the average flux density integrated over the band, $S_{\mathrm{w}}$, and the monochromatic flux density, $S_{0}$, at the central frequency, $\nu_{0}$, for two reasons. Firstly, most sources are better described by a power-law slope across the band than a simple linear slope. $S_{\mathrm{w}}$ will always exceed $S_{\mathrm{0}}$ for a source with a power-law slope. The magnitude of this effect increases with fractional bandwidth and for a source with an increasingly non-flat spectrum. The second cause of the discrepancy is the inverse noise-squared weighting applied to the 7.68~MHz sub-band mosaics: in practice, the noise in the 7.68~MHz sub-band mosaics decreases slightly with frequency, causing more weight to be assigned to higher frequency mosaics. For a source with $\alpha < 0$, where $\alpha$ is the spectral index ($S \propto \nu^{\alpha}$), these two effects go in opposite directions: $S_{\mathrm{w}}$ increases as a result of the power-law slope of sources across the band and decreases as a result of the weighting scheme adopted in the mosaicking.

For sources detected in each of the wide-band mosaics, we calculate the required flux density correction factor, $S_{0}/S_{\mathrm{w}}$. At any position in the mosaic,
\begin{equation}
S_{\mathrm{w}} = \Sigma_{i=1}^{N} w_{i} S_{0} \left(\frac{\nu_{i}}{\nu_{0}}\right)^{\alpha} \,,
\end{equation}
where $w_{i}$ is the weight assigned to the $i^{\mathrm{th}}$ sub-band, normalised such that $\Sigma_{i=1}^{N} w_{i} = 1.0$, $\nu_{i}$ is the central frequency of the $i^{\mathrm{th}}$ sub-band and $N$ is the number of 7.68~MHz sub-bands. The flux density correction factor is given by
\begin{equation}
\frac{S_{0}}{S_{\mathrm{w}}} = \left[\Sigma_{i=1}^{N} w_{i} \left(\frac{\nu_{i}}{\nu_{0}}\right)^{\alpha}\right]^{-1} \, .
\end{equation}

We produce simulated images of the flux density correction factor using the mosaicking software \textsc{swarp} \citep{bertin2002} assuming $\alpha = -0.8$, the typical spectral index of GLEAM sources between 76 and 227~MHz. Using these images we extract the correction factor for sources detected in each of the wide-band images. We find that the mean $\pm$ standard deviation of the correction factor in the 200, 154, 118 and 88~MHz mosaics is $1.000 \pm 0.009$, $1.003 \pm 0.001$, $1.007 \pm 0.002$ and $1.002 \pm 0.004$ respectively. Given the correction factors are very close to unity ($< 1$ per cent), we ignore them.

%typical of radio sources whose emission is dominated by optically thin synchrotron radiation.

%%%%%%%%%%%%%%%%%%%%%%%%%%%%%%%%%%%%%%%%%%%%%%%%%%%%%%%%%%%%%%%%%%%
\section{Determining the source counts}\label{Determining the source counts}

We measure the source counts at 200~MHz using the wide-band flux densities quoted in the GLEAM catalogue and at 154, 118 and 88~MHz using the catalogues compiled in Section~\ref{Source finding}. At each frequency, the vast majority of sources are point-like due to the large beam size. For unresolved sources, peak flux densities will be significantly more accurate than integrated flux densities at low signal-to-noise ratio (SNR). This is because more free parameters are required to measure an integrated flux density using Gaussian fitting. We note that peak flux densities are corrected for ionospheric smearing as outlined in Section~\ref{Effect of ionospheric smearing on the point spread function}. Therefore, in measuring the counts, we only use integrated flux densities for sources which are significantly resolved and use peak flux densities for the remaining sources. We distinguish between point-like and extended sources as described in Section~\ref{Classifying sources as point-like or extended}.

The rms noise varies substantially across the survey due to varying observational data quality and the presence of image artefacts originating from bright sources and the Galactic Plane. It increases at lower frequency and becomes less Gaussian as the classical confusion noise becomes more dominant. The counts must be corrected for both incompleteness and Eddington bias \citep{eddington1913} close to the survey detection limit. Incompleteness causes the counts to be underestimated close to the detection limit, 
while the Eddington bias makes it more likely for noise to scatter sources above the detection limit than to scatter them below it due to the steepness of source counts, consequently boosting the counts in the faintest bins. The magnitude of the Eddington bias only depends on the SNR and the source count slope \citep{hogg1998}.

The number of synthesised beams per source is often used as a measure of confusion as it indicates the typical separation of sources at the survey cut-off limit. The number of beams per source at each frequency is indicated in Table~\ref{tab:source_stats}. It is only 24 at the lowest frequency, indicating that the average separation between sources is $\sqrt{24} \approx 5$ beams. \cite{vernstrom2016} used simulated images to investigate the effect of confusion on the source-fitting accuracy for the source finders \textsc{aegean} and \textsc{obit} \citep{cotton2008}. Similar results were obtained for both source finders: sources separated by less than the beam size were fitted as a single source up to 95 per cent of the time, while the total flux density of the sources was, on average, conserved. Thus the effect of confusion is either to prevent a source from being detected or boost its flux density, which may, in turn, significantly bias the counts. In Section~\ref{Completeness and confusion}, we use Monte Carlo simulations to investigate the effect of incompleteness, Eddington bias and source blending on the counts.

%some (more?) discussion on confusion possibly affecting the counts with such a large beam - i.e. detecting and counting one bright source rather than multiple fainter sources near to each other. Is it a problem, if not, why not Ñ> something to do with source density/clustering at these flux densities.

%I did a fair bit of simulation tests regarding blending and source fitting in Vernstrom et al. (2016) where I put sources of varying flux densities (4 to 100 sigma) at set distances to each other (less than the beam size and greater than the beam size) in multiple resolutions and looked at the effect on the source finding. I found, using two different source finders, that even if both sources were 100 sigma, or 1 100 sigma and 1 10, the source finders didnÕt distinguish as two sources until the separation was greater than the beam size. When only one source is found/fit the total flux density of the two combined is ~conserved (==flux density of the one fitted source), but that would have an effect on the counts.

% Deep 3 GHz observations of the Lockman Hole with the VLA - source extraction and uncertainty analysis

Conversely, sources (i.e. physical entities associated with a host galaxy) of largest angular size may also be broken up into multiple components in GLEAM. In measuring the source counts, physically related components should be counted as a single source and their flux densities summed together. In Section~\ref{Complex sources}, we show that, given the large beam size, the source counts are well approximated as counts of components.

%%%%%%%%%%%%%%%%%%%%%%%%%%%%%%%%%%%%%%%%%%%%%%%%%%%%%%%%%%%%%%%%%%%
\subsection{Classifying sources as point-like or extended}\label{Classifying sources as point-like or extended}

We use the method described in \cite{franzen2015} to identify extended sources based on the ratio of integrated flux density, $S$, to peak flux density $S_{\mathrm{peak}}$. Assuming that the uncertainties on $S$ and $S_{\mathrm{peak}}$ ($\sigma_{S}$ and $\sigma_{S_{\mathrm{peak}}}$ respectively) are independent, to detect source extension at the 2$\sigma$ level, we require
\begin{equation}\label{eqn:indicator}
\ln\left(\frac{S}{S_{\mathrm{peak}}}\right) > 2 \sqrt{ \left(\frac{\sigma_{\mathrm{S}}}{S} \right)^{2} + \left(\frac{\sigma_{S_{\mathrm{peak}}}}{S_{\mathrm{peak}}} \right)^{2}} \ .
\end{equation}

We take $\sigma_{S_{\mathrm{peak}}}$ and $\sigma_{S}$ as the sum in quadrature of the Gaussian parameter fitting uncertainties returned by \textsc{aegean}, which accounts for the local noise, and the GLEAM internal flux density calibration error. The latter is estimated to be 2 per cent at $-72^{\circ} \leq \mathrm{Dec} < 18.5^{\circ}$ and 3 per cent at $\mathrm{Dec} < -72^{\circ}$ and $\mathrm{Dec} \geq 18.5^{\circ}$ \citep{hurleywalker2017}. For bright sources, where the 2 per cent calibration error dominates, $\frac{S}{S_{\mathrm{peak}}} > 1.06$ is considered to be extended.

\begin{figure}
\begin{center}
\includegraphics[scale=0.4, trim=0cm 0cm 0cm 0cm]{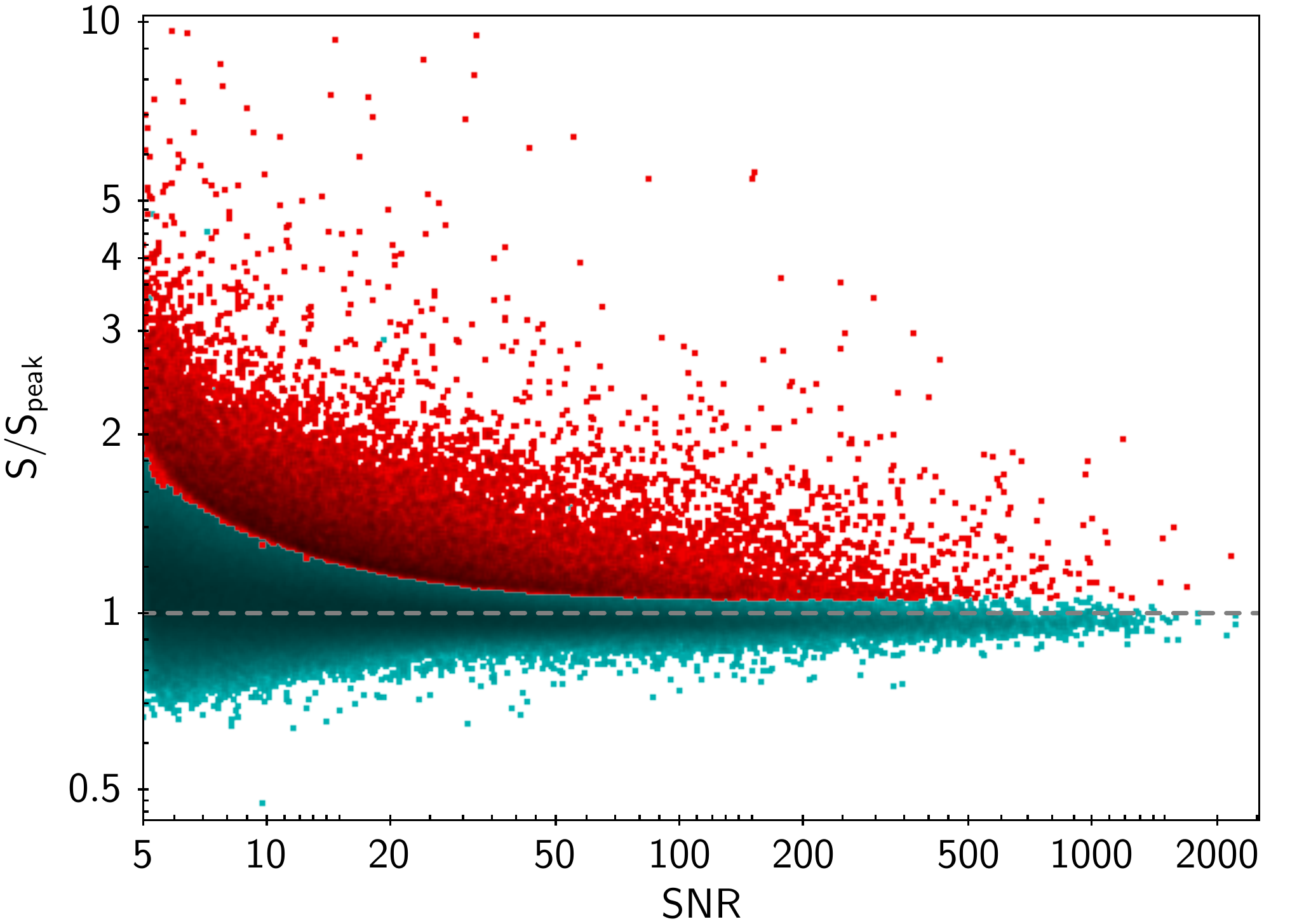}
\end{center}
\caption {$S/S_{\mathrm{peak}}$ as a function of SNR for all components detected at 200~MHz. The peak flux density values have been corrected for ionospheric smearing as described in Section~\ref{Effect of ionospheric smearing on the point spread function}. Components which are classified as point-like/extended are shown in turquoise/red.}
\label{fig:source_extension_white}
\end{figure}

Table~\ref{tab:source_stats} gives the fraction of sources classified as extended at each frequency. Fig.~\ref{fig:source_extension_white} shows $\frac{S}{S_{\mathrm{peak}}}$ as a function of SNR for all sources detected at 200~MHz. 7.3 per cent of sources are classified as extended at this frequency, where the beam size ($\approx 2.5$~arcmin) is smallest; these are highlighted in red.

\begin{figure}
\begin{center}
\includegraphics[scale=0.35, trim=2cm 0cm 0cm 0cm]{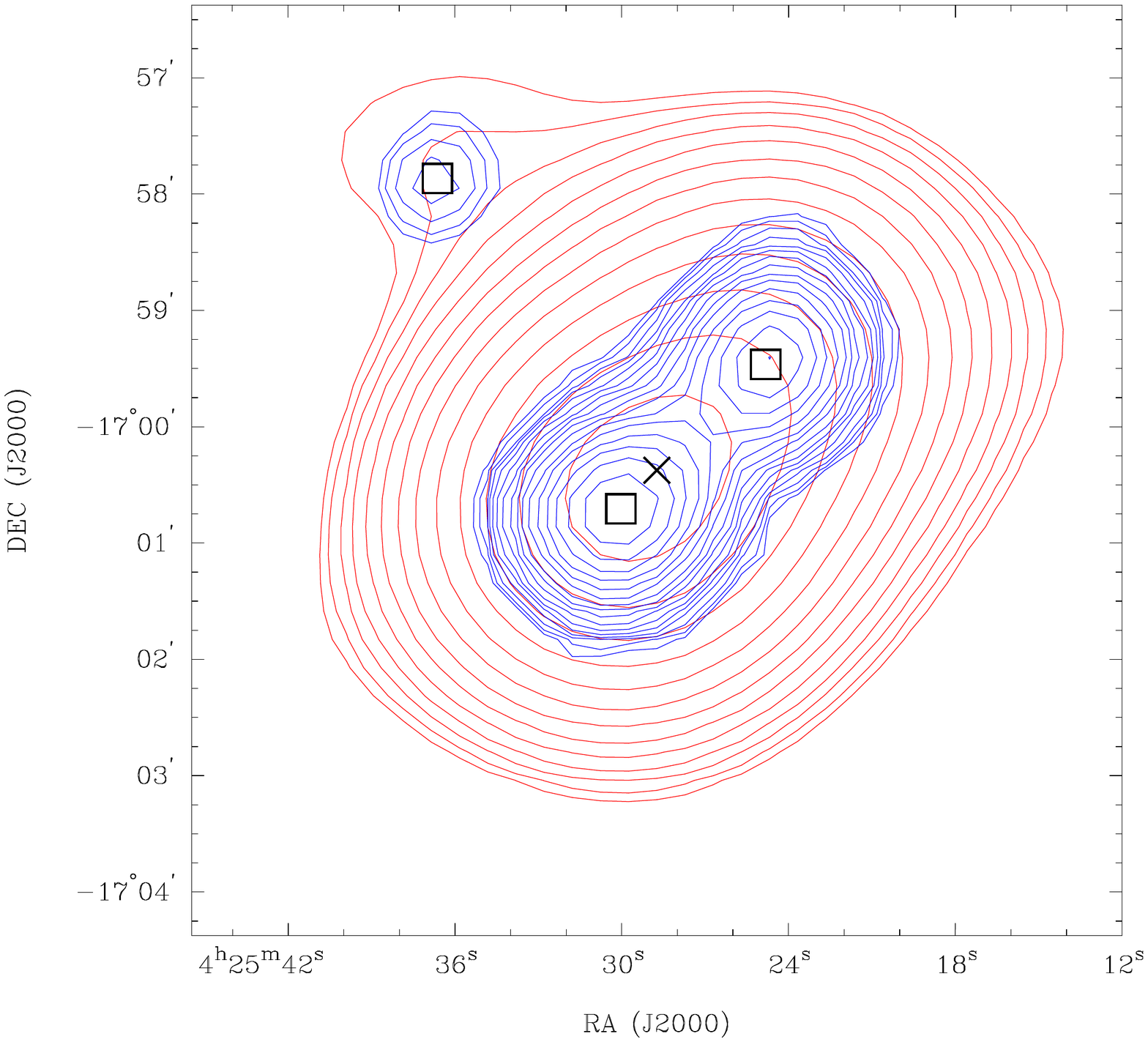}
\includegraphics[scale=0.35, trim=2cm 1cm 0cm 3cm]{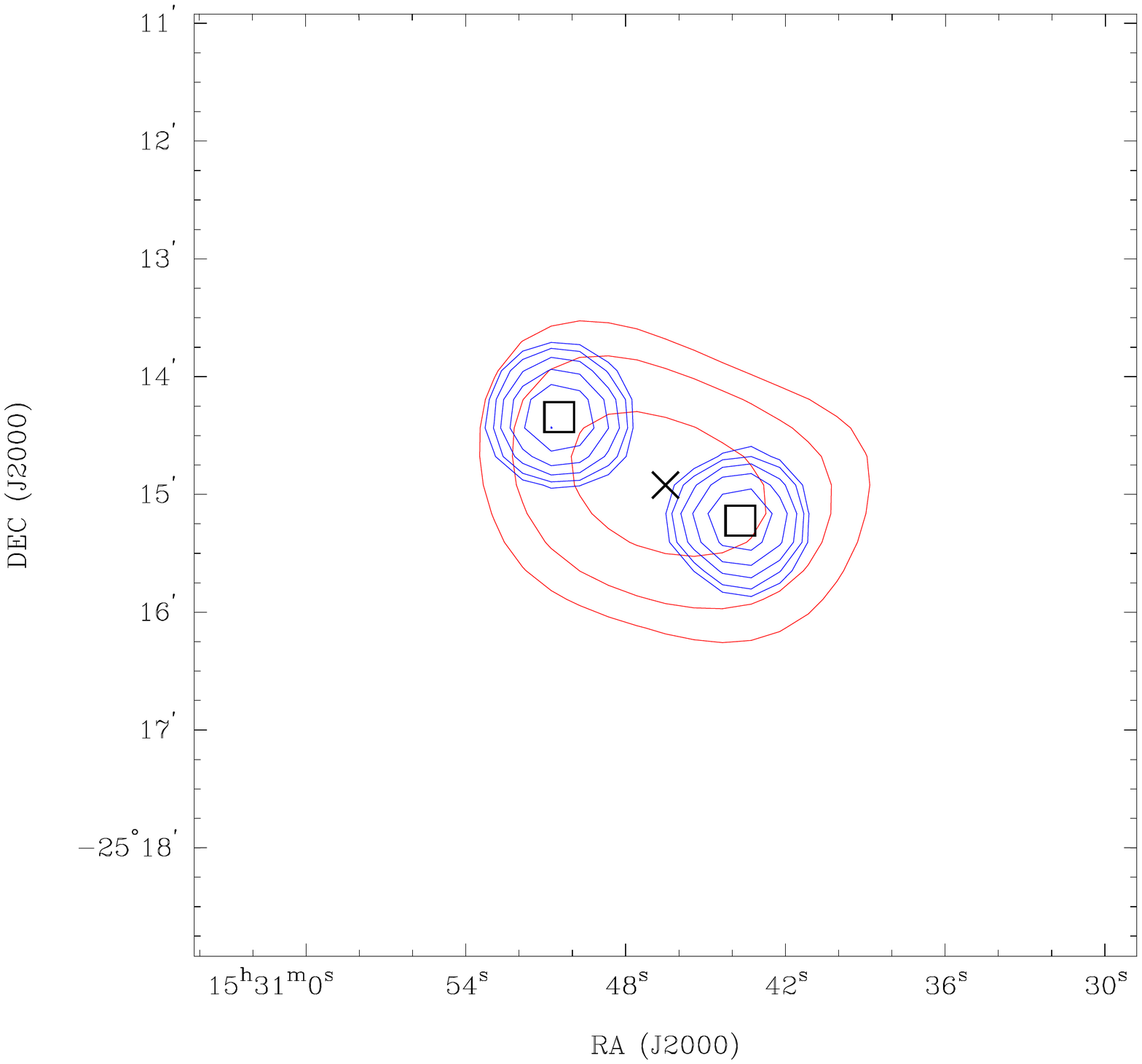}
\caption{An example of an extended GLEAM source associated with a resolved NVSS double (top) and with two NVSS components determined to be unrelated (bottom). Red (GLEAM) and blue (NVSS) contours are shown with the lowest contour level at 3$\sigma$; the contour levels increase at each level by a factor of $\sqrt{2}$. GLEAM and NVSS component positions are represented as crosses and squares respectively.}
\label{fig:gleam_and_nvss}
\end{center}
\end{figure}

Investigations using higher resolution (45~arcsec) radio images from the NRAO VLA Sky Survey \citep[NVSS;][]{condon1998} at 1.4~GHz show that a large fraction of resolved sources in GLEAM are, in fact, artefacts of source confusion or noise fluctuations: we randomly select 50 sources classified as extended at 200~MHz in the region of sky covered by NVSS, i.e. at $\mathrm{Dec} > -40^{\circ}$. We find that 39 of the sources are resolved into multiple components in NVSS. Of these 39 sources, only 16 are likely to be genuinely extended because the NVSS components have similar peak flux densities and there is extended emission linking the components; the remaining 23 sources probably appear extended as a result of source blending. An example of each of these cases is shown in Fig.~\ref{fig:gleam_and_nvss}.

%\cite{proctor2016} identify over 1600 candidate Giant Radio Sources (GRS) in NVSS using pattern recognition techniques. GRS are interpreted broadly to include any physically associated radio system with projected angular size $\geq 4$~arcmin. Such sources would appear extended in GLEAM.

%Since several arcmin seems quite large for source size (especially to have ~7 per cent this big), what are these sources? did you do any follow up in the images and with higher res images, ie TGSS? Is it like AGN with multiple components at high res but just slightly extended point source at 3arcmin? Can you say anything new or interesting about the physical sizes of sources from this (eg does 7 per cent match previous expectations for radios sources a few arcmin in size?)

%from NVSS only a few hundred GRGs identified, at most maybe 1 or 2 thousand (so barely a few percent).
%Proctor2016 --> identify over 1600 candidate GRS in NVSS using pattern recognition techniques
%GRS interpreted broadly to include any physically associated radio system with projected angular size > 4 arcmin
%Such sources would therefore appear extended in GLEAM

%%%%%%%%%%%%%%%%%%%%%%%%%%%%%%%%%%%%%%%%%%%%%%%%%%%%%%%%%%%%%%%%%%%
\subsection{Correcting the counts for incompleteness, Eddington bias and source blending}\label{Completeness and confusion}

We conduct Monte Carlo simulations to quantify the effect of incompleteness, Eddington bias and source blending on the counts. Our approach is to inject synthetic point sources with a range of flux densities into the wide-band images using \textsc{aeres} from the \textsc{aegean} package. We then use exactly the same source-finding procedure as described in Section~\ref{Source finding} to detect the simulated sources and measure their flux densities. The corrections to the counts as a function of flux density are obtained from the ratio of the injected count to the measured count of the simulated sources.

The major and minor axes of the simulated sources are set to $a_\mathrm{psf}$ and $b_\mathrm{psf}$ respectively, which are obtained from the PSF map at the relevant frequency. The simulated sources lie at random positions within region A but we set a minimum separation of 20~arcmin ($\approx 4$ times the beam size at the lowest frequency) between simulated sources to avoid them affecting each other. A simulated source may lie too close to a real ($>5\sigma$) source to be detected separately. In such situations, if the recovered source is closer to the simulated source than the real source, the simulated source is considered to be detected, otherwise not. Thus we account for source confusion in the counts in this analysis.

%If the source lies sufficiently close to a real source, it will either not be detected or its flux density will be boosted.

%This process can be repeated a large number of times with a range of injected flux densities. For practical reasons, we inject the sources into the images simultaneously; the simulated sources lie at random positions but we set a minimum separation of 20~arcmin ($\approx 4$ times the beam size at the lowest frequency) between simulated sources to avoid them affecting each other. 

It is important to ensure that the flux density distribution of the simulated sources is as realistic as possible and extends to well below the $5\sigma$ detection limit ($\gtrsim 50$~mJy/beam at 154~MHz). This is because the Eddington bias is dependent on the slope of the counts and causes the flux densities of sources with low SNRs to be biased high, boosting the number of sources detected in the faintest bins. The flux density distribution of the simulated sources at 154~MHz is based on the following source count model: above 33~mJy, we use a $3^{\mathrm{rd}}$ order polynomial fit to 154~MHz counts from a 12~hour pointed MWA observation of an EoR field, covering $570~\mathrm{deg}^{2}$ \citep{franzen2016}. Between 6 and 33~mJy, deep 153~MHz GMRT counts from \cite{williams2013} and \cite{intema2011} are well represented by a power law of slope $\gamma = 0.96$, where $S^{2.5} \frac{dN}{dS} = k S^{\gamma}$. We therefore set $\gamma = 0.96$ in this flux density range. A total of 40,000 flux densities ranging between 6~mJy and 15~Jy are drawn randomly from the source count model. We extrapolate the simulated source flux densities to 200, 118 and 88~MHz assuming $\alpha = -0.8$, as indicated by the typical spectral index seen in GLEAM.

%My approach is equivalent to adding a simulated source one at a time in the image, seeing if I detect the source and measuring its flux. It is possible that I donÕt detect a simulated source if it lies close enough to a real source (as explained in the text). I inject a realistic distribution of point sources (using information from deeper source counts) and see what flux distribution I recover. The fluxes of the simulated sources vary between 6 mJy and 15 Jy, but I only measure the counts above 80 mJy, which is  > 10 times brighter than the faintest simulated source. Of course it isnÕt practical to inject the sources one at a time in the image, so I inject a large number of sources in the image in one go, ensuring that the simulated sources don't overlap.

%The advantage of this approach is that I use the noise distribution from the real image in the simulations (I can account for the fact that it may not be Gaussian). I donÕt need to go into the trouble of subtracting the real sources from the image. The results of the simulations will also depend at some level on the flux distribution of the injected simulated sources, but I can use information from deeper source counts.

\begin{figure}
\begin{center}
\includegraphics[scale=0.35, trim=2.5cm 2cm 0cm 2cm]{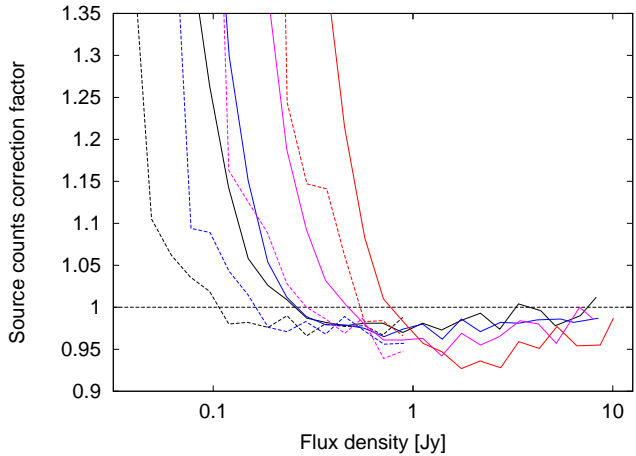}
\end{center}
\caption {Source count correction factor as a function of flux density at 200~MHz (black), 154~MHz (blue), 118~MHz (purple) and 88~MHz (red). The solid and dashed lines apply to regions A and B respectively. For clarity, the source count correction factor in region B is only shown below 1~Jy and error bars are not included.}
\label{fig:corr}
\end{figure}

The simulations are repeated 40 times to improve statistics. The solid lines in Fig.~\ref{fig:corr} show the mean source counts correction factor in region A, $c_{\mathrm{A}}$, in each of the wide-band images. The effects of both incompleteness and confusion are clearly evident. The sharp increase in the correction factor at low flux density is due to incompleteness. As expected, the survey becomes incomplete at a higher flux density in the lower frequency images. Source blending causes the correction factor to fall below 1.0 at higher flux densities. At 200~MHz, despite the large beam size of $\approx 2.5$~arcmin, the number of beams per source (49) is low enough for confusion not to strongly affect the counts, which are only overestimated by up to 2--3 per cent. As expected, the effect worsens at lower frequency due to the lower number of beams per source: at 88~MHz, the number of beams per source is 24 and the counts are overestimated by up to 7 per cent as a result of confusion.

%I mean one might automatically hear 3 arc minute beam and think well obviously theres multiple sources within 3 arc minutes, surely that must affect the fitting/finding/counting. But that may not be true or may not be significant depending on the source density at the given fluxes at these frequencies and the amount/level of expected source clustering. So I just thought a few sentences or something to that effect (as to why its not really problem) might be good (e.g. low source density at these flux densities and or low clustering).

From visual inspection of the rms noise maps, we identify areas within region A where the rms noise is well below average at zenith angles $\lessapprox 30$~deg, covering in total 6,516.2~$\mathrm{deg}^2$. The lines of RA and Dec bounding this region, hereafter referred to as region B, are given in Table~\ref{tab:regions}. The dashed lines show the correction factor in region B, $c_{\mathrm{B}}$. The counts start becoming incomplete at a flux density about twice as low as in region A at all frequencies. The counts are measured in region A in flux density bins where $c_{\mathrm{A}} \leq 1.2$. If $c_{\mathrm{A}} > 1.2$ and $c_{\mathrm{B}} \leq 1.2$, the counts are measured in region B. We do not measure the counts in bins where $c_{\mathrm{B}} > 1.2$ as the correction factor rises sharply with decreasing flux density in these bins and becomes unreliable.

\begin{table*}
\centering
\caption{Region B used to measure the source counts.}
\label{tab:regions}
\begin{tabular}{ c c c c c } 
\hline
RA range & Dec range & Area ($\mathrm{deg}^2$) \\
\hline
$10^{\mathrm{h}}00^{\mathrm{m}} < \alpha < 12^{\mathrm{h}}30^{\mathrm{m}}$ & $-40^{\circ} < \delta < -10^{\circ}$ & \multirow{2}{4em}{6,516.2} \\ 
$21^{\mathrm{h}}00^{\mathrm{m}} < \alpha < 06^{\mathrm{h}}15^{\mathrm{m}}$ & $-60^{\circ} < \delta < -10^{\circ}$ & \\ 
\hline
\end{tabular}
\end{table*}

%%%%%%%%%%%%%%%%%%%%%%%%%%%%%%%%%%%%%%%%%%%%%%%%%%%%%%%%%%%%%%%%%%%
\subsection{Complex sources}\label{Complex sources}

\begin{figure}
\begin{center}
\includegraphics[scale=0.33, trim=1cm 0cm 0cm 1cm]{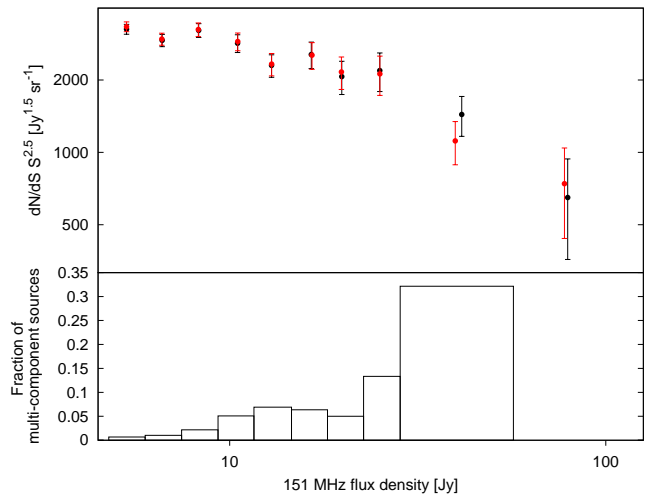}
\end{center}
\caption{Top: Euclidean normalised ($S^{2.5} \frac{dN}{dS}$) differential counts of the GLEAM 4~Jy sample at 151~MHz. The red circles show component counts while the black circles show counts for integrated sources. Bottom: fraction of multi-component sources in each flux density bin.}
\label{fig:counts_4Jy}
\end{figure}

We report counts of components rather than counts for integrated sources. The magnitude of the difference between the two will depend on the beam size and the intrinsic angular source size distribution. White et al., in prep., are analysing a subset of the GLEAM catalogue in detail to study the nature and evolution of the bright end of the low frequency population. The GLEAM 4~Jy sample is a statistically complete sample of 1845 sources with $S_{151 \mathrm{MHz}} > 4.0$~Jy, covering region A. Only 44 (2.4 per cent) of the sources are resolved into multiple components, where the beam size is $\approx 2.5$~arcmin. Multi-component sources are identified through visual inspection of higher resolution radio images from NVSS, the Sydney University Molonglo Sky Survey \citep[SUMSS;][]{mauch2007} and the Faint Images of the Radio Sky at Twenty Centimetres \citep[FIRST;][]{becker1995} survey. The likelihood of a source showing complex structure increases with flux density above 4~Jy due to the increasing fraction of objects at very low redshifts, as shown in the bottom panel of Fig.~\ref{fig:counts_4Jy}. No multi-component sources are detected in the highest flux density bin ($57-114$~Jy) but it only contains 5 sources, 3 of which are extended in GLEAM and resolved into multiple components in NVSS/SUMSS.

We use the GLEAM 4~Jy sample to measure both the source and component counts at $S_{151 \mathrm{MHz}} > 4.0$~Jy. We find that the component and source counts agree within the Poisson uncertainties, as shown in the top panel of Fig.~\ref{fig:counts_4Jy}, given the small fraction of sources which are resolved into multiple components. \cite{windhorst1990} found that, below $S_{1.4 \mathrm{GHz}} = 3$~Jy, the median angular size of radio galaxies, $\theta_{\mathrm{med}}$, decreases continuously towards fainter flux densities, with $\theta_{\mathrm{med}} \propto (S_{1.4~\mathrm{GHz}})^{0.3}$. Assuming that a similar relation holds at lower frequency, we expect our multi-frequency component counts to be a good approximation of the counts for integrated sources.

Finally, we note that the following bright, complex sources were peeled from the GLEAM data and subsequently lie outside region A: Hydra A, Pictor A, Hercules A, Virgo A, Crab, Cygnus A and Cassiopeia A. Centaurus A also lies outside region A. From measurements over 60--1400~MHz available via the NASA/IPAC Extragalactic Database (NED)\footnote{http://ned.ipac.caltech.edu/}, these sources are all brighter than 100~Jy at 200, 154, 118 and 88~MHz. Since our highest source count bin does not exceed 100~Jy at any of these frequencies, the exclusion of these sources does not bias our source count measurements.

\begin{figure*}
\begin{center}
\includegraphics[scale=0.55, trim=2cm 2.5cm 0cm 2.5cm]{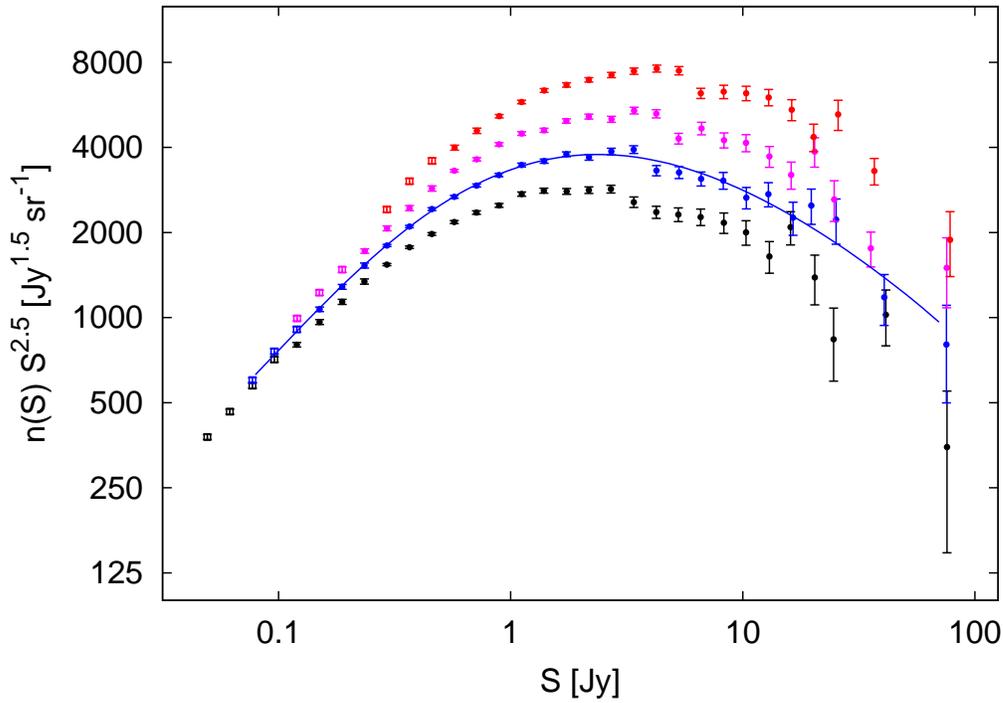}
\end{center}
\caption {Euclidean normalised differential counts at 200~MHz (black), 154~MHz (blue), 118~MHz (purple) and 88~MHz (red) from GLEAM. The different symbols distinguish between the areas used to derive the counts in the various flux density bins: the filled circles correspond to region A while the open squares correspond to region B. The solid blue line is a weighted least squares $5^{\mathrm{th}}$ order polyomial fit to the 154~MHz counts.}
\label{fig:counts_4freq}
\end{figure*}

%%%%%%%%%%%%%%%%%%%%%%%%%%%%%%%%%%%%%%%%%%%%%%%%%%%%%%%%%%%%%%%%%%%
\subsection{Analysis of the GLEAM source counts}\label{Results}

\begin{figure*}
\begin{center}
\includegraphics[scale=0.45, trim=0cm 0cm 0cm 2.5cm]{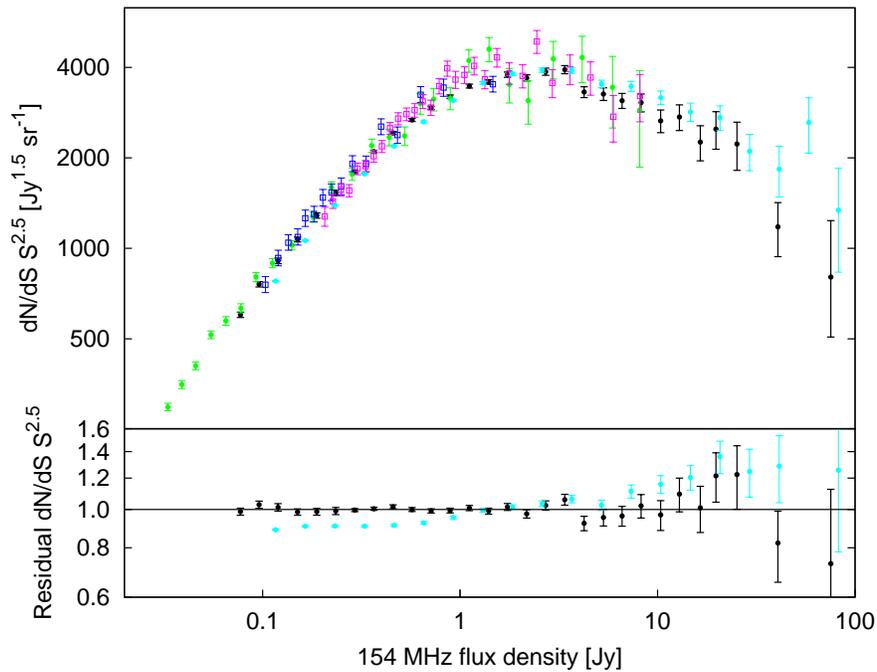}
\end{center}
\caption {Top: Euclidean normalised differential counts in the frequency range 150--154~MHz, extrapolated to 154~MHz assuming $\alpha = -0.8$. Black circles: this paper; green circles: \citet{franzen2016}; turquoise circles: \citet{intema2016}; purple squares: \citet{hales2007}; blue squares: \citet{mcgilchrist1990}. Bottom: the GLEAM and TGSS counts are normalised with respect to a polynomial fit to the GLEAM counts to highlight differences.}
\label{fig:counts_compare_literature_blue}
\end{figure*}

\begin{figure*}
\begin{center}
\includegraphics[scale=0.45, trim=0cm 0cm 0cm 1cm]{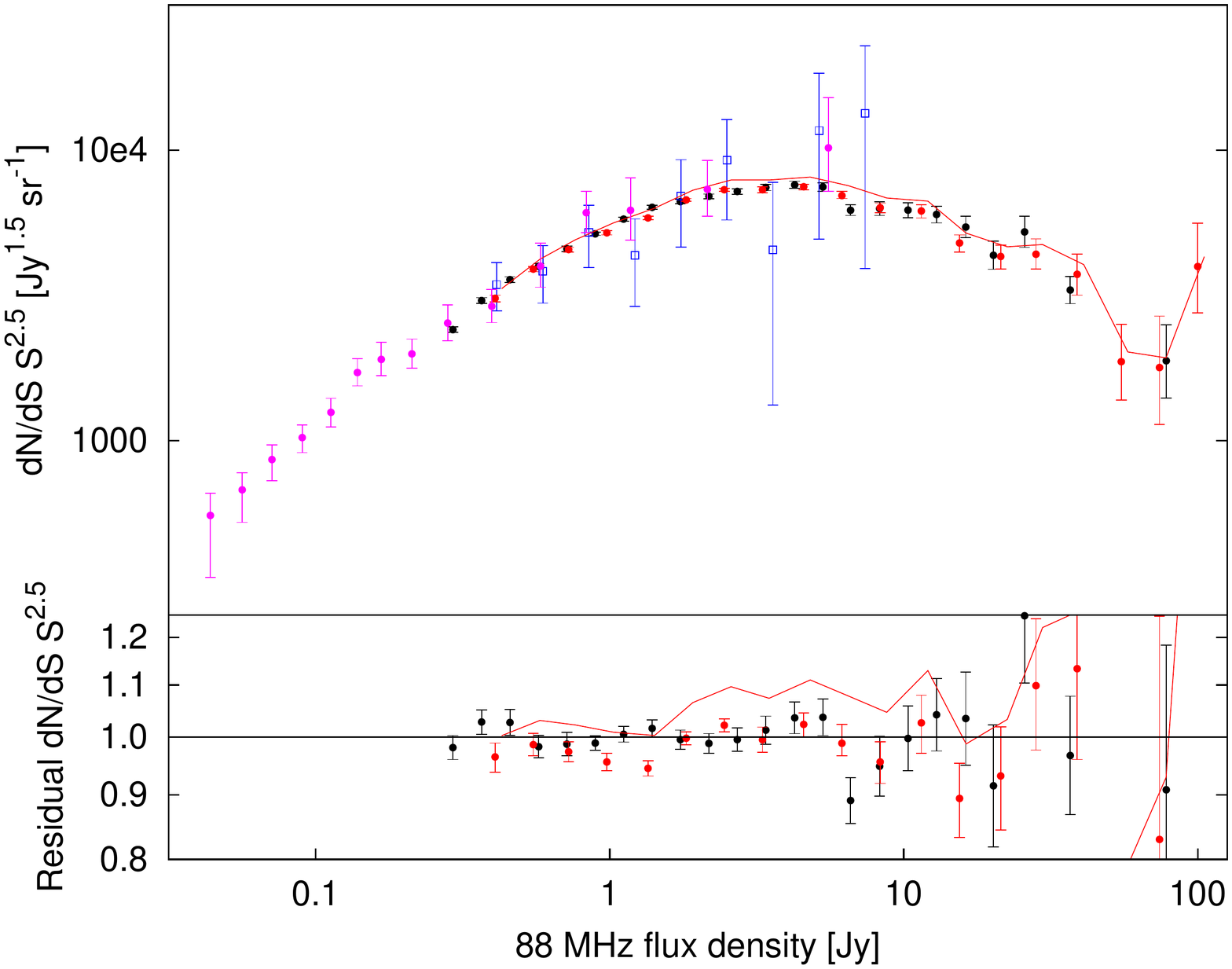}
\end{center}
\caption {Top: Euclidean normalised differential counts in the frequency range 62--93.75~MHz, extrapolated to 88~MHz assuming $\alpha = -0.8$. Black circles: this paper; red circles: \citet{lane2014}; blue squares: \citet{zheng2016}; purple circles: \citet{van_weeren2014}. The red line displays the 74~MHz counts by \citet{lane2014} scaled with $\alpha = -0.5$. Bottom: the GLEAM and VLSSr counts are normalised with respect to a polynomial fit to the GLEAM counts to highlight differences.}
\label{fig:counts_compare_literature_red}
\end{figure*}

The corrected GLEAM differential source counts are shown in Fig.~\ref{fig:counts_4freq}, while the source count data are provided in Table~\ref{tab:gleam_source_counts}. Uncertainties on the counts are propagated from Poisson errors on the number of sources per bin and the errors on the correction factors derived in Section~\ref{Completeness and confusion}. The Poisson error on $N$ is approximated as $\sqrt{N}$ in all bins with $N \geq 20$. In bins with $N < 20$, we use approximate expressions for 84 per cent confidence upper and lower limits based on Poisson statistics by \citet{gehrels1986}. 

The bulge due to source evolution is clearly evident at all four frequencies given the large areal sky coverage and the range of flux densities sampled. A detailed comparison of the shape of the multi-frequency counts is undertaken in Section~\ref{Scaling of counts with frequency}.

In Fig.~\ref{fig:counts_compare_literature_blue}, we compare the GLEAM counts with other counts in the literature at a similar frequency covering more than $100~\mathrm{deg}^{2}$: the 154~MHz counts by \cite{franzen2016}, 7C counts at 151~MHz by \cite{mcgilchrist1990} and \cite{hales2007} and TGSS First Alternative Data Release (ADR1) counts at 150~MHz by \cite{intema2016}. The 7C and TGSS counts are extrapolated to 154~MHz assuming $\alpha = -0.8$. 

The GLEAM counts are generally in excellent agreement with the other counts. We note that GLEAM and TGSS are on different flux density scales, with TGSS on the scale of \cite{scaife2012}. There is, however, a flux density dependent offset between the GLEAM and TGSS counts. While the ratio of TGSS to GLEAM counts lies close to 1.0 at a few Jy, it decreases to $\approx 0.9$ below $\sim 1$~Jy. This is consistent with a $\approx 6$ per cent decrease in the mean ratio of TGSS to GLEAM flux densities below $\sim 1$~Jy and may be due to missing low surface brightness emission in TGSS. The TGSS observations have a far less centrally concentrated $uv$ coverage than the GLEAM observations. At 154~MHz, GLEAM has a resolution of $\approx 3$~arcmin while TGSS has a resolution of 25 by $25/\cos(\delta-19^{\circ})$~arcsec.

Source counts below 100~MHz are comparatively sparse. In Fig.~\ref{fig:counts_compare_literature_red}, we compare the 88~MHz GLEAM counts with the VLSSr counts at 74~MHz, placed on the \cite{baars1977} flux density scale \citep{lane2014}; 62~MHz counts from LOFAR observations of the 3C295 and Bo{\"o}tes fields, covering $36~\mathrm{deg}^{2}$ \citep{van_weeren2014}; and 93.75~MHz counts from a 12 hour pointed observation with the 21 Centimetre Array (CMA) of a $25~\mathrm{deg}^{2}$ region of sky coincident with the North Celestial Pole \citep{zheng2016}. The GLEAM counts, which cover the largest area of sky, show good agreement with the other counts extrapolated to 88~MHz with $\alpha = -0.8$. Below $\sim 1$~Jy, the GLEAM counts lie very slightly (2--3 per cent) above the VLSSr counts but this is sensitive to the spectral index used in the extrapolation. We note that VLSSr has a resolution of 75~arcsec as compared to the GLEAM resolution of $\approx 5$~arcmin at 88~MHz.

%%%%%%%%%%%%%%%%%%%%%%%%%%%%%%%%%%%%%%%%%%%%%%%%%%%%%%%%%%%%%%%%%%%
\section{Investigating changes in the source count shape with frequency}\label{Scaling of counts with frequency}

In this section, we analyse any change in the shape of the GLEAM counts with frequency and the dependence of the spectral index on flux density and frequency. We also show that the behaviour of the counts is broadly consistent with the typical spectra of sources across the MWA band. 

\begin{figure}
 \begin{center}
  \includegraphics[scale=0.33, trim=1cm 0cm 0cm 5cm]{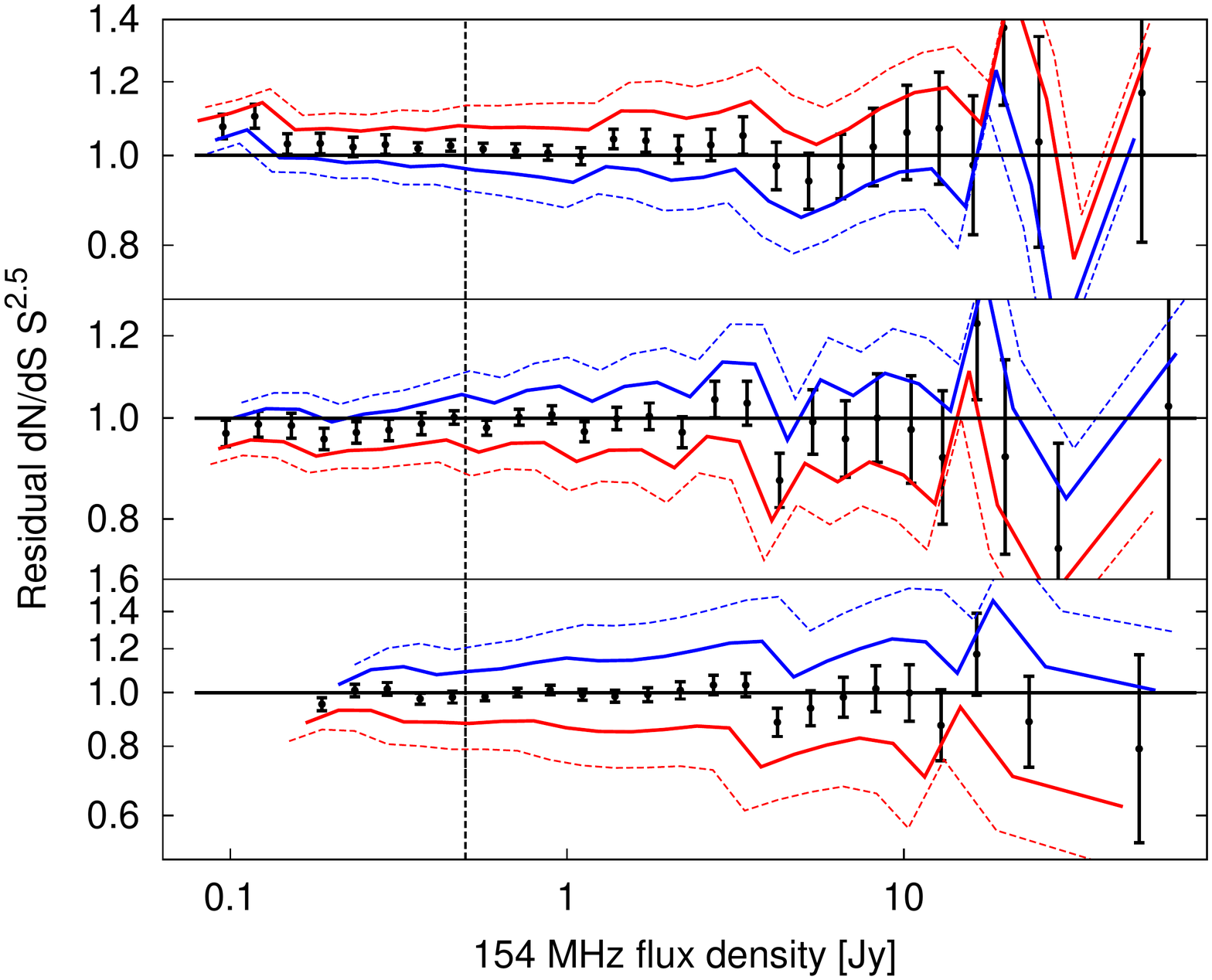}
  \caption{Top: the 200~MHz GLEAM counts are extrapolated to 154~MHz and divided by a polynomial fit to the 154~MHz GLEAM counts to highlight differences. The spectral index used in the extrapolation is --0.4 (dashed blue line), --0.6 (solid blue line), --0.8 (black circles),  --1.0 (solid red line) and --1.2 (dashed red line). The 200~MHz counts are replaced by the 118 and 88~MHz counts in the central and bottom panels respectively. At $S_{154~\mathrm{MHz}} > 0.5$~Jy (dashed vertical line), $\alpha \approx -0.8$ provides a good match between the counts.}
  \label{fig:counts_residual}
 \end{center}
\end{figure}

The solid line in Fig.~\ref{fig:counts_4freq} is a weighted least squares $5^{\mathrm{th}}$ order polynomial fit to the GLEAM 154~MHz counts. We extrapolate the 200, 118 and 88~MHz GLEAM counts to 154~MHz assuming various spectral indices and divide the extrapolated counts by the 154~MHz source count fit calculated above, as shown in Fig.~\ref{fig:counts_residual}.

We find that, at $S_{154~\mathrm{MHz}} \gtrsim 0.5$~Jy, there is no significant change in the shape of the counts at the four frequencies. We calculate the value of $\alpha$ which minimises the $\chi^{2}$ difference between the counts at each of the three pairs of frequencies. When computing $\chi^{2}$, we exclude the region of the 154~MHz source count fit below 0.5~Jy. For example, for the 154--200~MHz source count pair, 
\begin{equation}
\chi^{2} = \Sigma_{i=1}^{N} w_{i} \left[  n_{i,200} - y n_{154} \left(\frac{S_{i,200}}{x}\right)   \right]^{2} \,,
\end{equation}
where
\begin{eqnarray}
w_{i} =
\left\{
\begin{array}{ll}
\left[ \sigma_{n_{i,200}}^{2} + \sigma_{n_{154} (S_{i,200}/x)}^{2} \right]^{-1}    & \mathrm{if~} \frac{S_{i,200}}{x} > 0.5~\mathrm{Jy,} \\
0 & \mathrm{otherwise.}
\end{array}
\right.
\end{eqnarray}
$n_{i,200}$ is the Euclidean normalised source count in the $i^{\mathrm{th}}$ bin at 200~MHz, $\sigma_{n_{i,200}}$ is the error on $n_{i,200}$, $n_{154} \left(\frac{S_{i,200}}{x}\right)$ is the 154~MHz source count fit above evaluated at $\frac{S_{i,200}}{x}$, $S_{i,200}$ is the central flux density of the $i^{\mathrm{th}}$ bin at 200~MHz, $x = (200/154)^{\alpha}$ and $y = x^{1.5}$. For the 154--200, 154--118 and 154--88~MHz source count pairs, $\chi^{2}$ is minimised with $\alpha =$ --0.75, --0.77 and --0.79 respectively. Thus there is no strong dependence of the spectral index on frequency.

At $S_{154~\mathrm{MHz}} < 0.5$~Jy, it becomes hard to discriminate between different spectral indices given the steep slope of the counts. There is, however, tentative evidence that a flatter spectral index provides a better match between the 154--200 and 154--118~MHz source count pairs.

\begin{figure}
 \begin{center}
  \includegraphics[scale=0.35, trim=2cm 2cm 0cm 2cm]{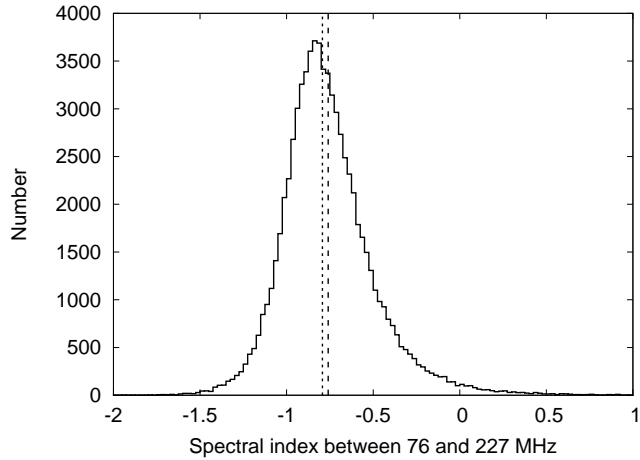}
  \caption{Spectral index distribution between 76 and 227~MHz for sources with $S_{200~\mathrm{MHz}} > 60$~mJy in region B of the GLEAM catalogue. The vertical dotted and dashed lines show the median and mean values of --0.79 and --0.76 respectively.}
  \label{fig:alpha_dist}
 \end{center}
\end{figure}

\cite{hurleywalker2017} calculated the 76--227~MHz spectral indices of sources in the GLEAM catalogue using the 7.68~MHz sub-band flux densities. For the spectral index of a source to be quoted in the catalogue, the source must have a positive flux density in each of the 20 sub-bands (this is not always the case at low SNR) and the spectrum must be well fit by a power-law. From the completeness maps presented in \citeauthor{hurleywalker2017}, in region B, the GLEAM catalogue is 90 per cent complete at $S_{200~\mathrm{MHz}} = 60$~mJy. Of the 84,003 sources with $S_{200~\mathrm{MHz}} > 60$~mJy in region B, 75,905 (90.4 per cent) have measured spectral indices in the GLEAM catalogue. Fig.~\ref{fig:alpha_dist} shows the spectral index distribution for these sources. The distribution is roughly symmetric about the median value of --0.79 but there is a positive tail which extends to $\alpha \approx 0.5$.

%\textcolor{red}{
%\cite{condon1984b} measured spectral index distributions between 1.4 and 5~GHz of sources selected at 1.4~GHz; these distributions have a full width at half-maximum (FWHM) of $\approx 0.4$. The spectral index distribution shown in Fig.~\ref{fig:alpha_dist} also has a FWHM of $\approx 0.4$. A number of other low-frequency spectral index distributions, including those presented in \cite{marvil2015}, \cite{heald2015} and \cite{hardcastle2016}, have a significantly larger FWHM. These distributions are probably broader due to the effect of larger flux-density measurement errors.
%}

% () \citep[e.g.][]{condon1984b}.

\begin{figure}
 \begin{center}
  \includegraphics[scale=0.32, trim=0cm 0cm 0cm 1cm]{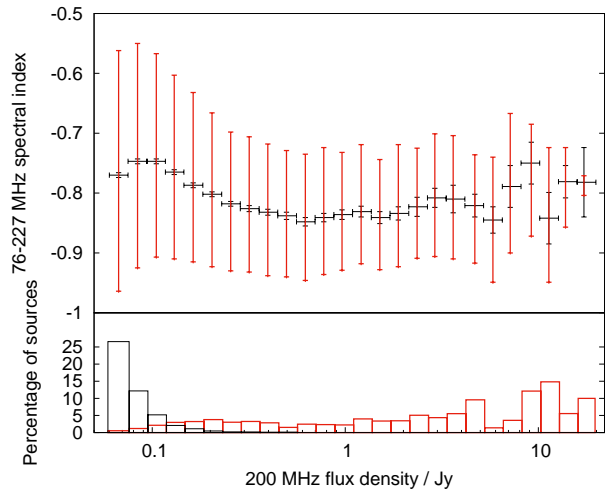}
  \caption{Top: the black data points show the median 76--227~MHz spectral index as a function of flux density; the error bars are standard errors of the median. The red bars extend from the first to the third quartile. Bottom: percentage of sources which have no measured spectral indices in the GLEAM catalogue because they are not well fit by a power-law (red) or because they have a negative flux density in at least one of the 7.68~MHz sub-bands (black).}
  \label{fig:stats_alpha}
 \end{center}
\end{figure}

The top panel of Fig.~\ref{fig:stats_alpha} shows the median spectral index, $\alpha_{\mathrm{med}}$, as a function of $S_{200~\mathrm{MHz}}$. Sources which are missing from the spectral index sample because they are not well fit by a power-law are represented by the red histogram in the bottom panel. These sources include compact-steep spectrum (CSS) sources with a peak in their spectra across the MWA band, hypothesized to be the precursors to massive radio galaxies, and are studied in detail in \cite{callingham2017}.

Above 0.5~Jy, we find that there is no significant change in the median spectral index, $\alpha_{\mathrm{med}}$, with flux density, whereas $\alpha_{\mathrm{med}}$ flattens from $\approx -0.85$ to $\approx -0.75$ between 0.5 and 0.1~Jy. We caution that $\alpha_{\mathrm{med}}$ is biased towards steep values below 0.1~Jy. Indeed, a substantial fraction of sources have no measured spectral indices in bins below 0.1~Jy because they do not have positive flux densities in all sub-bands; the negative flux densities mostly occur in lower frequency sub-bands due to the low SNR (see black histogram in bottom panel of Fig.~\ref{fig:stats_alpha}). This probably explains the steepening in $\alpha_{\mathrm{med}}$ with decreasing flux density below 0.1 Jy.

Spectral flattening towards lower frequencies is expected for some sources due to absorption effects including synchrotron self-absorption and thermal absorption of a synchrotron power-law component. Spectral ageing, which causes the spectrum to steepen towards higher frequencies, may introduce additional curvature in the source spectrum.

%Hotspots in the steep-spectrum radio lobes of powerful ($L_{178~\mathrm{MHz}} > 10^{27}~\mathrm{W}~\mathrm{Hz}^{-1}~\mathrm{sr}^{-1}$) radio galaxies often have low-frequency ($\lesssim 300$~MHz) spectra that flatten towards lower frequency, possibly due to a low-energy cutoff in the energy distribution of the relativistic particles within the hotspots (see e.g. \cite{leahy1989}). Depending on the fractional contribution of the hotspots to the total radio flux density, this may significantly affect the integrated spectrum. The effect is more likely to be seen at high redshift where Inverse Compton scattering results in a substantial dimming of the diffuse synchrotron emission from the lobes.

Given the weak dependence of the median redshift of radio galaxies on flux density \citep[see e.g.][]{condon1993}, the flux-density range 0.1--0.5~Jy is expected to correspond to the least-luminous radio galaxies. By studying a number of complete samples of radio sources at frequencies close to 151~MHz with good coverage of the luminosity-redshift plane, \cite{blundell1999} found an anti-correlation between the rest-frame spectral index at low frequency and the source luminosity. This correlation is understood to arise through the steepening of the injection spectrum of particles by radiative losses in the enhanced magnetic fields of the hotspots of sources with more powerful jets. It is possible that the spectral flattening observed for GLEAM sources in this flux density range also results from this effect.

At $S_{154~\mathrm{MHz}} > 0.5$~Jy, we find no evidence of any flattening in the average spectral index with decreasing frequency. Van Weeren et al. (2014) measured source counts at 34, 46 and 62~MHz down to 136, 72 and 51~mJy respectively, from LOFAR observations of the 3C295 and Bo{\"o}tes fields, covering a few tens of square degrees (their 62~MHz counts are displayed in Fig.~\ref{fig:counts_compare_literature_red} of this paper). They found that (1) the 62~MHz counts are in good agreement with 153~MHz GMRT and 74~MHz VLA counts, scaling with $\alpha = -0.7$; (2) the 34~MHz counts fall significantly below the extrapolated counts from 74 and 153~MHz with $\alpha = -0.7$. Instead, $\alpha = -0.5$ provides a better match to the 34~MHz counts.

\begin{figure}
\begin{center}
\includegraphics[scale=0.35, trim=2cm 0cm 0cm 0cm]{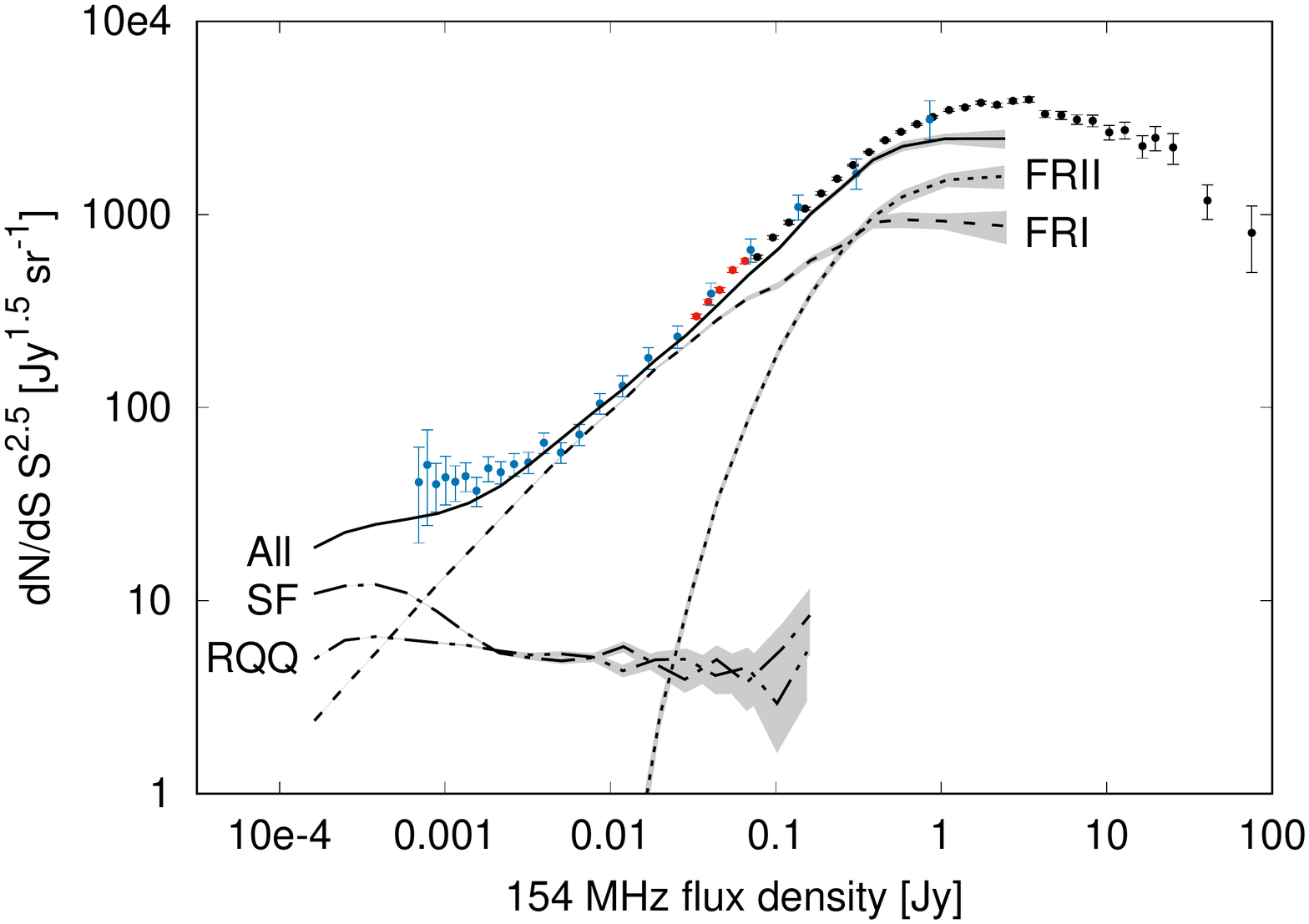}
\includegraphics[scale=0.35, trim=2cm 1cm 0cm 3cm]{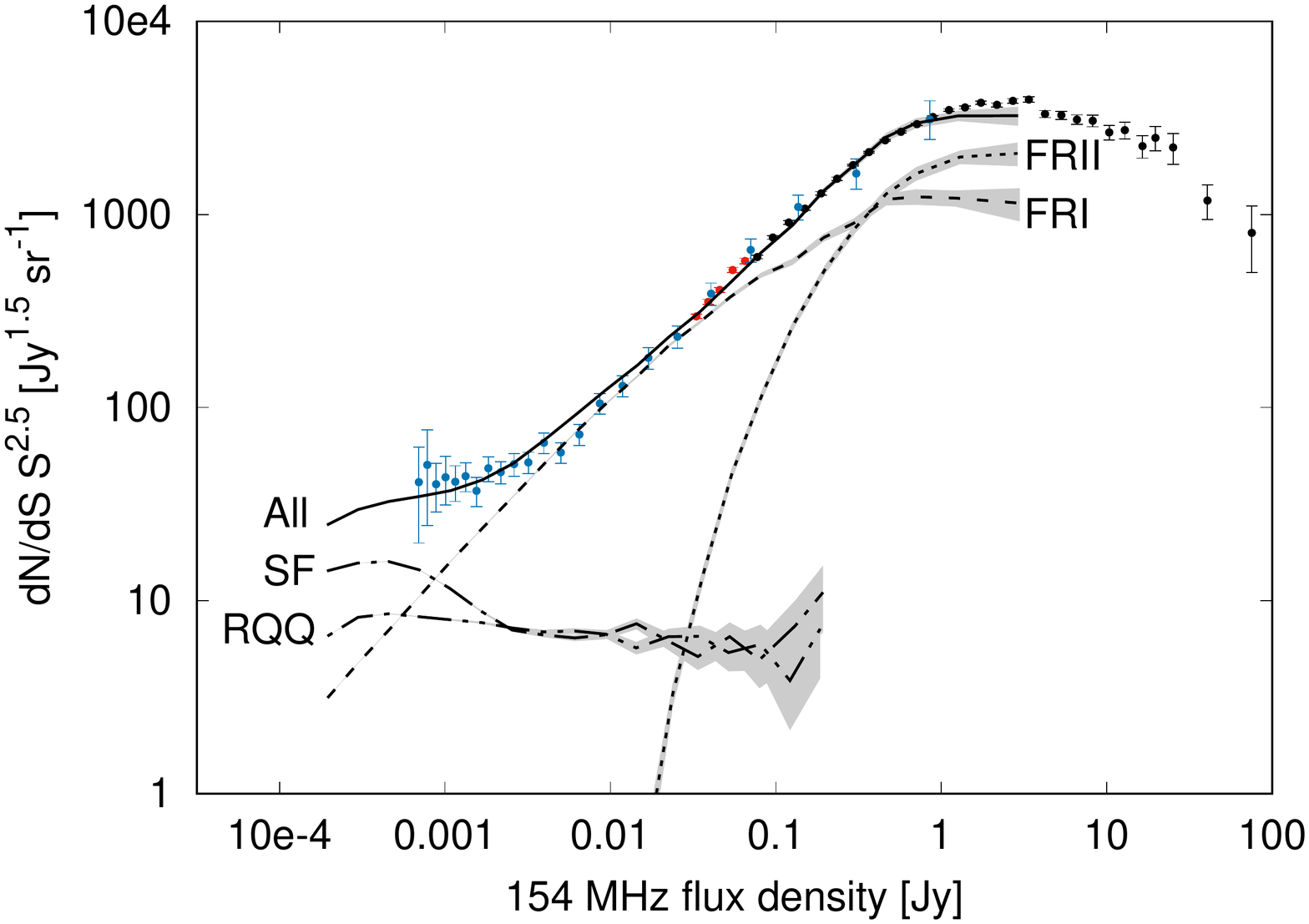}
\caption{Top: the data points show the counts from this paper (black), MWA counts from \cite{franzen2016} (red) and LOFAR counts from \cite{williams2016} (blue) at 154~MHz. These are compared with the SKADS simulations by \citet{wilman2008}, including contributions from FRI and FRII sources, star-forming galaxies and radio-quiet AGN. The 151~MHz SKADS model count is extrapolated to 154~MHz assuming $\alpha = -0.8$. The shaded area indicates the $1\sigma$ errors. Bottom: same as above except that the simulated flux densities in the model are multiplied by 1.2 to obtain a better fit to the data.}
\label{fig:skads_comparison}
\end{center}
\end{figure}

\begin{figure}
\begin{center}
\includegraphics[scale=0.45, trim=3cm 1cm 0cm 0cm]{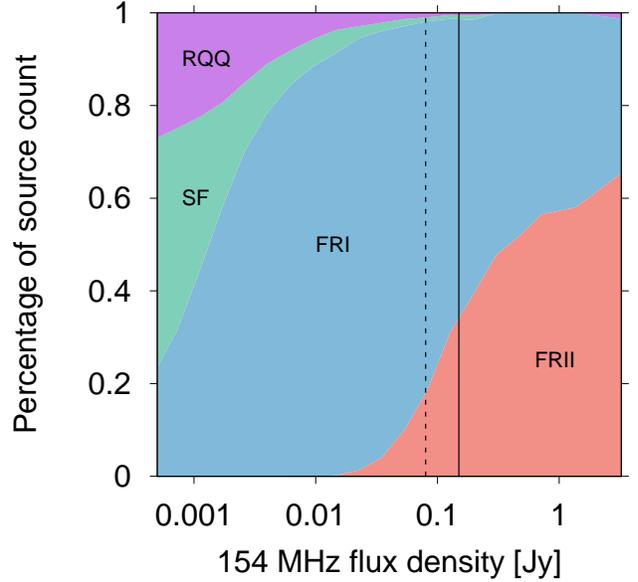}
\caption{The population mix at 154~MHz as predicted by the SKADS model after multiplying the simulated flux densities by 1.2. For reference, the 154~MHz GLEAM counts are measured above 150~mJy (solid line) and 80~mJy (dashed line) in regions A and B respectively.}
\label{fig:pop_division}
\end{center}
\end{figure}

%%%%%%%%%%%%%%%%%%%%%%%%%%%%%%%%%%%%%%%%%%%%%%%%%%%%%%%%%%%%%%%%%%%
\section{Comparison with SKADS Simulated Skies}\label{Comparison with model}

The SKADS model by \cite{wilman2008} gives radio flux densities at 151~MHz, 610~MHz, 1.4~GHz, 4.86~GHz and 18~GHz, down to 10~nJy, in a sky area of $20 \times 20~\mathrm{deg}^{2}$, and includes four distinct source types: FRI and FRII sources, radio-quiet AGN and star-forming galaxies. We compare observed counts at 154~MHz covering over 5 orders of magnitude in flux density with the source count prediction from the simulated database. We use the 154~MHz GLEAM counts, the deeper MWA EoR counts in the flux density range 30--75~mJy and 150~MHz LOFAR counts by \cite{williams2016}, extrapolated to 154~MHz with $\alpha = -0.8$.

In the top panel of Fig.~\ref{fig:skads_comparison}, we see that the 151~MHz SKADS model lies within the scatter of the observations except at $S \gtrsim 50$~mJy, where it increasingly underpredicts the measured counts with flux density. The GLEAM counts provide a very stringent test above this flux density given their high precision. The model underpredicts the number of sources by $\approx 50$ per cent by $\approx 2$~Jy. Since the model only covers $400~\mathrm{deg}^{2}$, the source population is too poorly sampled above this flux density to perform a precise comparison.

\cite{mauch2013} compared 325~MHz counts from a GMRT survey of the \textit{Herschel}-ATLAS/GAMA fields with the SKADS model and found a similar result, albeit to a lower significance. They determined the 325~MHz simulated flux density by calculating the power-law spectral index between 151 and 610~MHz. Their measured counts, which sample the flux density range 10--200~mJy, tend to lie slightly above the simulated counts above $S_{325~\mathrm{MHz}} \approx 50$~mJy.

We find that the model is statistically in much better agreement with the data at high flux density after multiplying the simulated flux densities by 1.2, as shown in the bottom panel of Fig.~\ref{fig:skads_comparison}. The fit is also somewhat improved at the low flux density end sampled by LOFAR although the data points have larger error bars making it harder to assess the model's accuracy.

 \citeauthor{mauch2013} suggest that the simulated flux densities at low frequency could be too low as a result of excessive spectral curvature implemented in the model. However, it is difficult to see how this is possible: radio-loud AGN dominate the source population at $S_{154~\mathrm{MHz}} > 50$~mJy in the model. The overwhelming majority of these sources have power-law spectra between 154~MHz and 1.4~GHz, as the emission is lobe-dominated. 
 
%We also find that the 1.4~GHz SKADS model count lies significantly below observed 1.4~GHz source counts compiled by \cite{hopkins2003} at $S_{1.4~\mathrm{GHz}} > 10$~mJy.
  
 % source counts from the 3CRR \citep{laing1983}, 6C \citep{hales1988} and 7C \citep{mcgilchrist1990} surveys.
% Laing, Riley & Longair (1983)
% 3CRR: S_178 > 10 Jy, Dec > 10 deg and |b| > 10 deg
% 6C: S_151 > 190 mJy, 2030 deg^2
% 7C: S_151 > 80 mJy, 473 deg^2
 
%\textcolor{red}{
%At the bright end, the model is based on a compilation of source counts at 151~MHz by \cite{willott2001}. Despite this, the model also lies significantly below these counts. We note, however, that the discrepancy between the model and the GLEAM counts is stronger due to the smaller Poisson errors on the GLEAM counts. The model is also based on the 151~MHz luminosity function of high-luminosity radio galaxies by \citeauthor{willott2001}. They chose to fit a Schechter luminosity function, whose exponential high-luminosity cutoff is likely too sharp to describe radio galaxies.
 %}
 
At the bright end, the model is based on a compilation of source counts at 151~MHz by \cite{willott2001}. The GLEAM counts provide much tighter constraints. The model is also based on the 151~MHz luminosity function of high-luminosity radio galaxies by \citeauthor{willott2001}. They chose to fit a Schechter luminosity function, whose exponential high-luminosity cutoff is likely too sharp to describe radio galaxies.

Fig.~\ref{fig:pop_division} shows the fraction of each source type as a function of $S_{154~\mathrm{MHz}}$ as predicted by the SKADS model, after rescaling the simulated flux densities. According to the model, FRII sources are dominant above $\sim 500$~mJy, FRI sources in the flux density range $\sim 1-500$~mJy and star-forming galaxies below $\sim 1$~mJy.

 %%%%%%%%%%%%%%%%%%%%%%%%%%%%%%%%%%%%%%%%%%%%%%%%%%%%%%%%%%%%%%%%%%%
\section{Noise and confusion properties of GLEAM mosaics}\label{Noise properties of GLEAM images}

\begin{figure}
\begin{center}
\includegraphics[scale=0.35, trim=3cm 2cm 0cm 0cm]{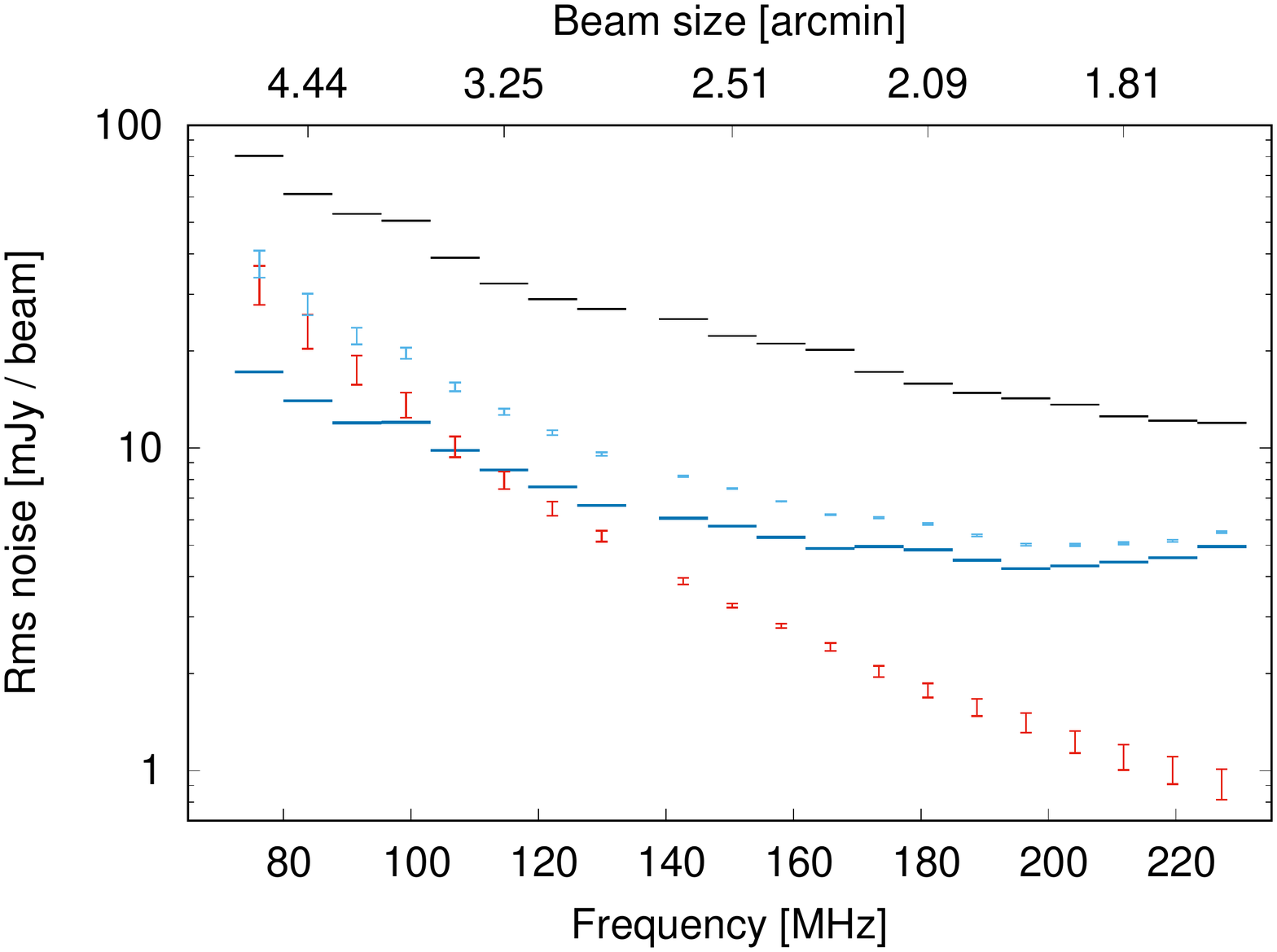}
\includegraphics[scale=0.35, trim=3cm 2cm 0cm 0cm]{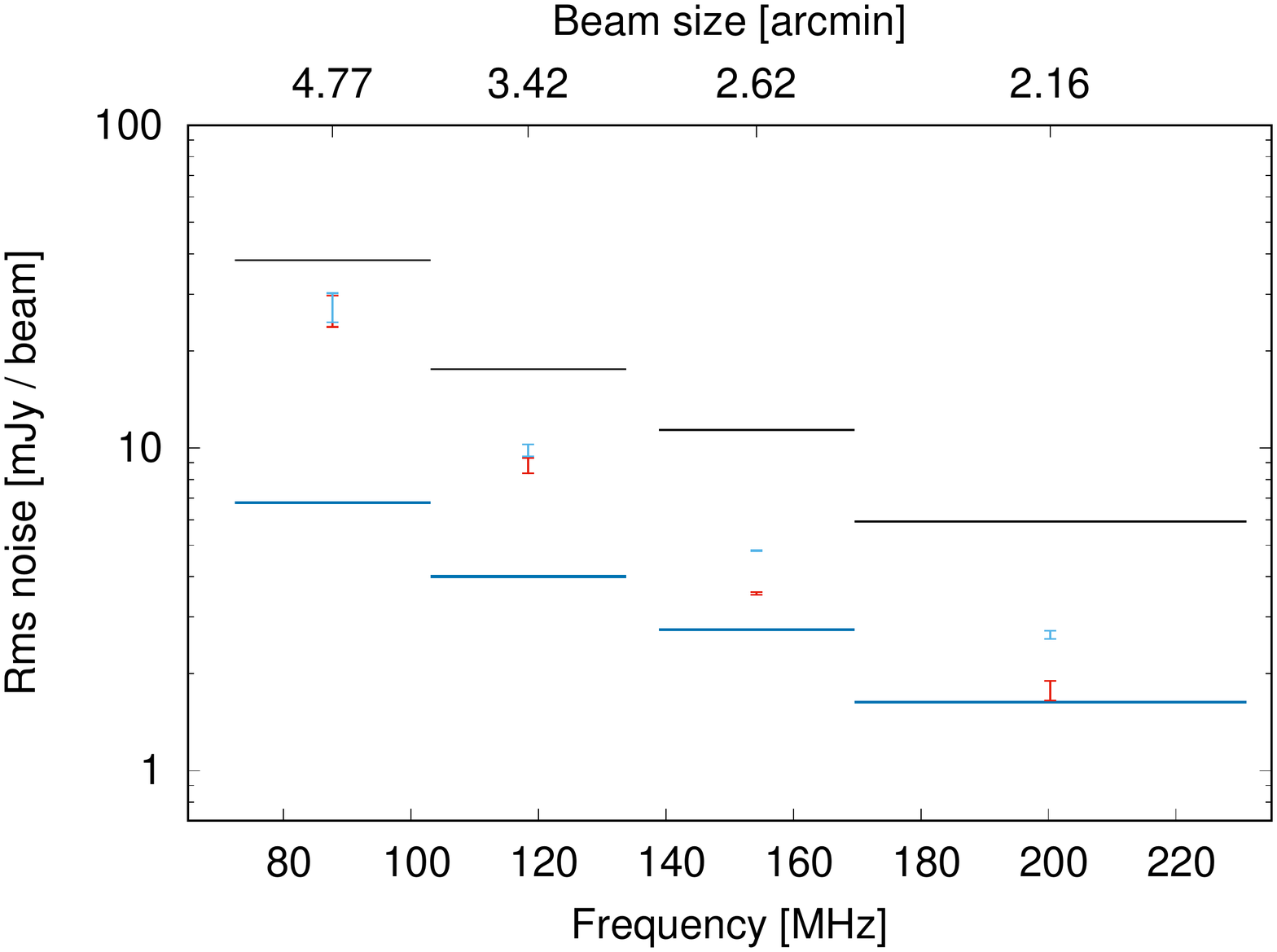}
\caption{Top: rms noise in the narrow-band mosaics in a region within 8.5~deg from CDFS (black horizontal bars), expected thermal noise sensitivity from Stokes $V$ mosaics (blue horizontal bars), range of classical confusion noise estimates (red) and range of theoretical noise limits (turquoise points). The approximate beam size is shown on the top. Bottom: same as above in the wide-band mosaics.}
\label{fig:noise_vs_freq}
\end{center}
\end{figure}

Fig.~\ref{fig:noise_vs_freq} shows the mean rms noise, measured using \textsc{bane}, in the narrow- and wide-band mosaics in a circular region within 8.5~deg of the \textit{Chandra} Deep Field-South (CDFS) at J2000 $\alpha = 03^{\mathrm{h}}30^{\mathrm{m}}$, $\delta = -28^{\circ}00'$, hereafter referred to as region C; this region lies close to zenith (i.e. at $\delta = -26.7^\circ$) and 55~deg from the Galactic Plane.

We derive the expected thermal noise in this cold region of extragalactic sky. We then use our knowledge of the low-frequency source counts below the flux densities sampled by GLEAM to derive the theoretical noise limit, accounting for both the thermal noise and classical confusion, and compare it with the measured rms noise.

%%%%%%%%%%%%%%%%%%%%%%%%%%%%%%%%%%%%%%%%%%%%%%%%%%%%%%%%%%%%%%%%%%%
\subsection{Estimating the thermal noise}\label{Estimating the thermal noise}

Since no circular polarisation is expected from extragalactic sources, Stokes $V$ images should provide a good measure of the thermal noise. We download all narrow-band, uniformly-weighted Stokes $V$ snapshot images contributing to region C from the GLEAM Data Centre\footnote{http://mwa-web.icrar.org/gleam/q/form}, originating from four different declination strips ($-13^\circ$, $-27^\circ$, $-40^\circ$ and $-55^\circ$). We verify that the rms noise in Stokes $V$ images from the Dec $-27^\circ$ strip is in good agreement with the theoretical prediction.

The naturally-weighted, point-source sensitivity of the MWA, in Jy/beam, is given by
\begin{equation}
\sigma_{\mathrm{t}} = \frac{2 k_{\mathrm{B}} T}{A_{\mathrm{eff}} \epsilon_{\mathrm{c}} } \sqrt{\frac{1}{\tau B n_{\mathrm{p}} N (N-1)}} \, ,
\label{eq:theoretical_sensitivity}
\end{equation}
where $k_{\mathrm{B}}$ is the Boltzmann constant, $T$ the system temperature in K, $A_{\mathrm{eff}}$ the effective area of each antenna tile in $\mathrm{m}^2$, $N$ the number of antenna tiles, $\epsilon_{\mathrm{c}}$ the correlator efficiency, $\tau$ the integration time in seconds, $B$ the bandwidth in Hz and $n_{\mathrm{p}}$ the number of polarisations \citep{tingay2013}. 

\begin{figure}
\begin{center}
\includegraphics[scale=0.35, trim=3cm 2cm 0cm 0cm]{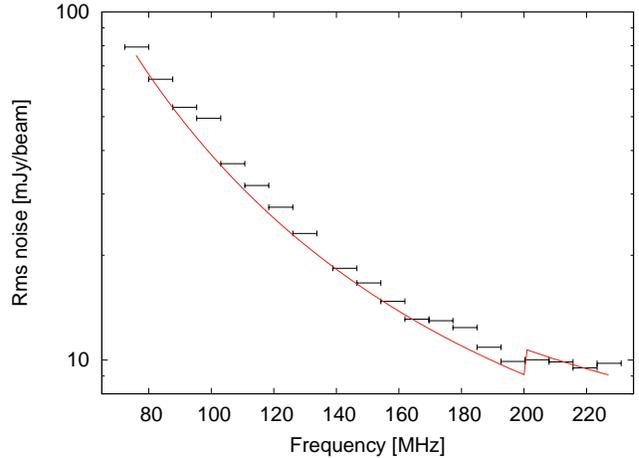}
\end{center}
\caption{Horizontal bars: rms noise in uniformly-weighted Stokes $V$ snapshot images with a bandwidth of 7.68~MHz, centred within a few deg from J2000 $\alpha = 03^{\mathrm{h}}30^{\mathrm{m}}$, $\delta = -28^{\circ}00'$. Red curve: theoretical noise prediction using equation~\ref{eq:theoretical_sensitivity}.}
\label{fig:thermal_noise_vs_freq}
\end{figure}

The system temperature is given by $T = T_{\mathrm{sky}} + T_{\mathrm{rec}}$, where $T_{\mathrm{sky}}$ is the sky temperature and $T_{\mathrm{rec}}$ the receiver temperature. \cite{wayth2015} present measurements of the average sky temperature for pointings at different declinations and LSTs at multiple GLEAM frequencies. From this information, we obtain $T_{\mathrm{sky}} \approx 228~\mathrm{K}~(\nu/150~\mathrm{MHz})^{-2.53}$ at the location of the CDFS. Following \cite{wayth2015}, we set $T_{\mathrm{rec}} = 50$~K except at $\nu > 200$~MHz, where we set $T_{\mathrm{rec}} = 80$~K; laboratory measurements by Sutinjo et al., in preparation, indicate that $T_{\mathrm{rec}} \approx 80$~K at $\nu > 200$~MHz. We set $B = 0.75 \times 7.68$~MHz given a 25 per cent reduction in the bandwidth due to flagged edge channels. We set the remaining parameters as follows: $A_{\mathrm{eff}} = 21.5~\mathrm{m}^2$, $N = 128$, $\epsilon_{\mathrm{c}} = 1.0$, $\tau = 2$~min and $n_{\mathrm{p}} = 2$. We also account for a 2.1-fold loss in sensitivity due to uniform weighting \citep{wayth2015}. We find that the theoretical prediction agrees within 25 per cent with the Stokes $V$ noise measurements across the entire frequency range (see Fig.~\ref{fig:thermal_noise_vs_freq}).

We combine all the Stokes $V$ snapshot images to produce narrow- and wide-band Stokes $V$ mosaics, following the procedure described in \cite{hurleywalker2017} for Stokes $I$. We measure the mean rms noise in region C of each Stokes $V$ mosaic. The blue horizontal bars in Fig.~\ref{fig:noise_vs_freq} show our thermal noise estimates for the narrow- and wide-band mosaics.

%%%%%%%%%%%%%%%%%%%%%%%%%%%%%%%%%%%%%%%%%%%%%%%%%%%%%%%%%%%%%%%%%%%
\subsection{Estimating the theoretical noise limit}\label{Estimating the theoretical noise limit}

Given a source count model and beam size, we use the method of probability of deflection \citep{scheuer1957} to derive the exact shape of the source $P(D)$ distribution, $P_{\mathrm{c}}(D)$, that is the probability distribution of pixel values resulting from all sources present in the image. We then estimate the rms classical confusion noise, $\sigma_{\mathrm{c}}$, from the core width of this distribution.

A detailed explanation of the equations used to derive the $P_{\mathrm{c}}(D)$ can be found in \cite{vernstrom2014}. Briefly, we calculate the mean number of pixels per steradian with observed intensities between $x$ and $x+dx$,
\begin{equation}
R(x)~dx = \int_{\Omega} \frac{dN}{dS} \left( \frac{x}{B(\theta,\phi)} \right) B(\theta,\phi)^{-1}~d\Omega~dx \, ,
\end{equation}
where $dN/dS$ is the differential source count and $x = S B(\theta,\phi)$ is the image response to a point source of flux density $S$ at a point in the synthesised beam where the relative gain is $B(\theta,\phi)$. The predicted $P_{\mathrm{c}}(D$ distribution is then computed from the Fourier Transform of $R(x)$, such that
\begin{equation}
P(D) = \mathcal{F}^{-1} [p(\omega)] \, ,
\end{equation}
where
\begin{equation}
p(\omega) = \exp \left[\int_{0}^{\infty} R(x)~\exp(i \omega x)~dx - \int_{0}^{\infty} R(x)~dx \right] \, .
\end{equation}

%\footnotesize
%\begin{equation}
%P(D) = \mathcal{F}^{-1} \left[ \exp \left(\int_{0}^{\infty} R(x)~\exp(i \omega x)~dx - \int_{0}^{\infty} R(x)~dx \right) \right] \, .
%\end{equation}
%\normalsize

\begin{figure}
\begin{center}
\includegraphics[scale=0.35, trim=2.6cm 2cm 0cm 0cm]{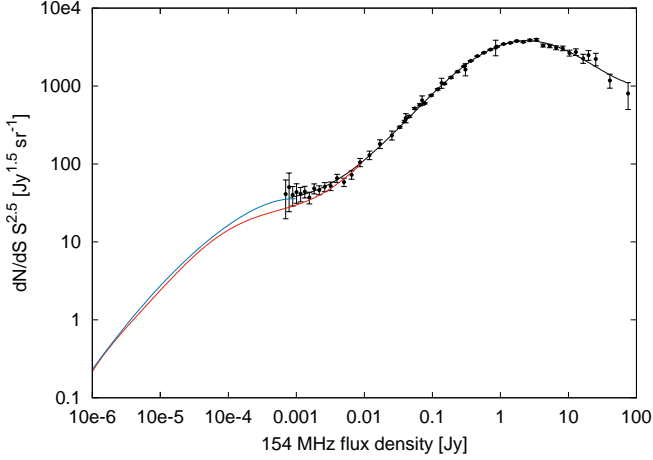}
\end{center}
\caption{The data points show the 154~MHz counts from this paper, \citet{franzen2016} and \citet{williams2016}. The black curve is a polynomial fit to these counts. The red curve shows the 151~MHz SKADS model count \citep{wilman2008} while the blue curve shows the 151~MHz SKADS model count, applying a flux density scaling factor of 1.2.}
\label{fig:model_counts}
\end{figure}

The black curve in Fig~\ref{fig:model_counts} is a weighted least squares $5^{\mathrm{th}}$ order polynomial fit to the 154~MHz GLEAM counts and the 150~MHz counts by \cite{williams2016}, extrapolated to 154~MHz with $\alpha = -0.8$. The polynomial fit is given by
\begin{equation}
\log_{10}\left(S^{2.5} \frac{dN}{dS}\right) = \sum_{i=0}^{5} a_{i} [\log_{10}(S)]^{i} \mathrm{,}
\end{equation}
where $a_{0} = 3.52$, $a_{1} = 0.307$, $a_{2} = -0.388$, $a_{3} = -0.0404$, $a_{4} = 0.0351$ and $a_{5} = 0.00600$. The fit is valid over the flux density range 1~mJy--75~Jy.

Since no 154~MHz source count data are available below $\approx 1~\mathrm{mJy}$, we use the 151~MHz SKADS model count after multiplying the simulated flux densities by 1.2 (see blue curve in Fig~\ref{fig:model_counts}). We choose to apply this flux density scaling factor as the model is then in better agreement with the observed counts above 1~mJy, as shown in Section~\ref{Comparison with model}. At 1~mJy, there is minimal discontinuity between the rescaled SKADS model and the above polynomial fit to the observed counts. Our preferred model, source count model A, consists of our polynomial fit to the observed counts above 1~mJy and the rescaled SKADS model below 1~mJy.

Below a few mJy, the LOFAR counts have relatively large uncertainties and the 151~MHz SKADS model, displayed as the red curve in Fig~\ref{fig:model_counts}, lies significantly below the LOFAR counts. There is minimal discontinuity between our polynomial fit to the observed counts and the SKADS model at 10~mJy. We therefore consider a second model, source count model B, consisting of the polynomial fit above 10~mJy and the SKADS model below 10~mJy.

In Section~\ref{Scaling of counts with frequency}, we showed that a spectral index scaling of $\approx -0.8$ provides a good match between the GLEAM counts at $S_{154~\mathrm{MHz}} > 0.5$~Jy. It is not clear whether this continues to be the case at lower flux densities. We extrapolate the models to other frequencies with $\alpha = -0.6$, --0.8 and --1.0 in order to gauge the effect of spectral indices flatter and steeper than --0.8 on $\sigma_{\mathrm{c}}$.

In calculating $P_{\mathrm{c}}(D)$, we assume that the beam is a circular Gaussian with a full width at half-maximum (FWHM) $\theta = \sqrt{ a_{\mathrm{psf, mean}} b_{\mathrm{psf, mean}} }$, where $a_{\mathrm{psf, mean}}$ and $b_{\mathrm{psf, mean}}$ are the mean values of $a_{\mathrm{psf}}$ and $b_{\mathrm{psf}}$ in region C of the PSF map, respectively. This accounts for the increase in area of the PSF, resulting from ionospheric smearing.

\begin{figure}
\begin{center}
\includegraphics[scale=0.45, trim=1cm 6cm 0cm 0cm]{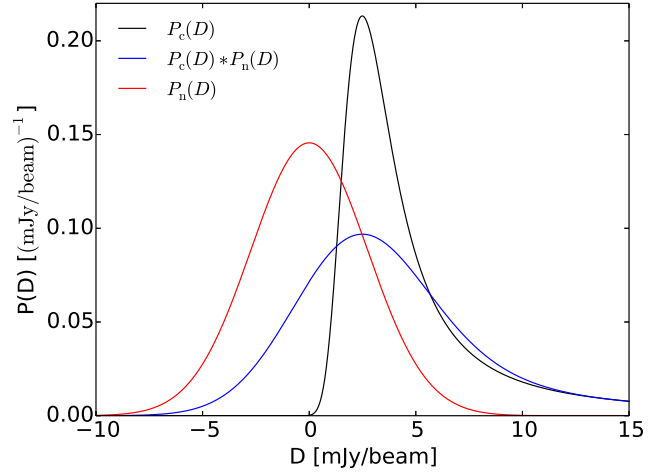}
\caption{Source $P(D)$ distribution (black curve), thermal noise distribution (red curve) and source $P(D)$ distribution convolved with thermal noise distribution (blue curve) in region C of the 139--170~MHz GLEAM mosaic. The source $P(D)$ distribution was derived using source count model A.}
\label{fig:pofd_examples}
\end{center}
\end{figure}

The black curve in Fig.~\ref{fig:pofd_examples} shows the $P_{\mathrm{c}}(D)$ distribution that we derive in the wide-band image at 139--170~MHz using source count model A, where $\theta = 2.6$~arcmin. The width of the distribution is measured by dividing the interquartile range by 1.349, i.e. the rms for a Gaussian distribution, obtaining $\sigma_{\mathrm{c}} = 3.6$~mJy/beam.

To account for the thermal noise, $\sigma_{\mathrm{t}}$, $P_{\mathrm{c}}(D)$ must be convolved with the thermal noise distribution, $P_{\mathrm{n}}(D)$, represented as a Gaussian with rms $\sigma_{\mathrm{t}}$. The convolution of $P_{\mathrm{c}}(D)$ with $P_{\mathrm{n}}(D)$ can be expressed as
\begin{equation}
P_{\mathrm{c}}(D) * P_{\mathrm{n}}(D) = \mathcal{F}^{-1} \left[p(\omega)~\exp \left(\frac{-\sigma_{\mathrm{t}}^{2} \omega^{2}}{2} \right) \right] \, .
\end{equation}

Our thermal noise estimate in region C of the 139--170~MHz mosaic is 2.7~mJy/beam. The red curve in Fig.~\ref{fig:pofd_examples} is a Gaussian centred on zero with a standard deviation of 2.7~mJy/beam, representing $P_{\mathrm{n}}(D)$, while the blue curve is the convolution of $P_{\mathrm{c}}(D)$ with $P_{\mathrm{n}}(D)$. The blue curve has a core width of 4.8~mJy/beam, and we take this to be the theoretical noise limit, $\sigma_{\mathrm{lim}}$.

We follow this procedure to derive $\sigma_{\mathrm{c}}$ and $\sigma_{\mathrm{lim}}$ for the narrow- and wide-band mosaics at all frequencies. We derive $\sigma_{\mathrm{c}}$ and $\sigma_{\mathrm{lim}}$ using both 154~MHz source count models and $\alpha = -0.6$, --0.8 and --1.0 to extrapolate the models to other frequencies. The range of $\sigma_{\mathrm{c}}$ and $\sigma_{\mathrm{lim}}$ values are displayed in Fig.~\ref{fig:noise_vs_freq}. We find that, at 154~MHz, $\sigma_{\mathrm{c}}$ changes by no more than 3 per cent depending on the source count model adopted. Varying the spectral index has a greater effect on $\sigma_{\mathrm{c}}$ at the upper and lower ends of the GLEAM frequency range.

%%%%%%%%%%%%%%%%%%%%%%%%%%%%%%%%%%%%%%%%%%%%%%%%%%%%%%%%%%%%%%%%%%%
\subsection{Excess background noise}\label{Excess background noise}

Fig.~\ref{fig:noise_vs_freq} reveals that the rms noise is a factor of $\approx 2-3$ higher than $\sigma_{\mathrm{lim}}$ in the narrow-band mosaics. The rms noise is a factor of $\approx 2$ higher than $\sigma_{\mathrm{lim}}$ in the wide-band mosaics at the highest 3 frequencies, while it is only $\approx 25$ per cent higher than $\sigma_{\mathrm{lim}}$ at the lowest frequency. The lowest frequency wide-band mosaic is limited by classical confusion since $\sigma_{\mathrm{c}}$ is a factor of $\approx 4$ higher than $\sigma_{\mathrm{t}}$.

%%%%%%%%%%%%%%%%%%%%%%%%%%%%%%%%%%%%%%%%%%%%%%%%%%%%%%%%%%%%%%%%%%%
\section{Origin of excess background noise in GLEAM images}\label{Origin of excess background noise in GLEAM images}

\begin{figure*}
\begin{center}
\includegraphics[scale=0.9,angle=270, trim=6.5cm 5cm 2.5cm 0cm]{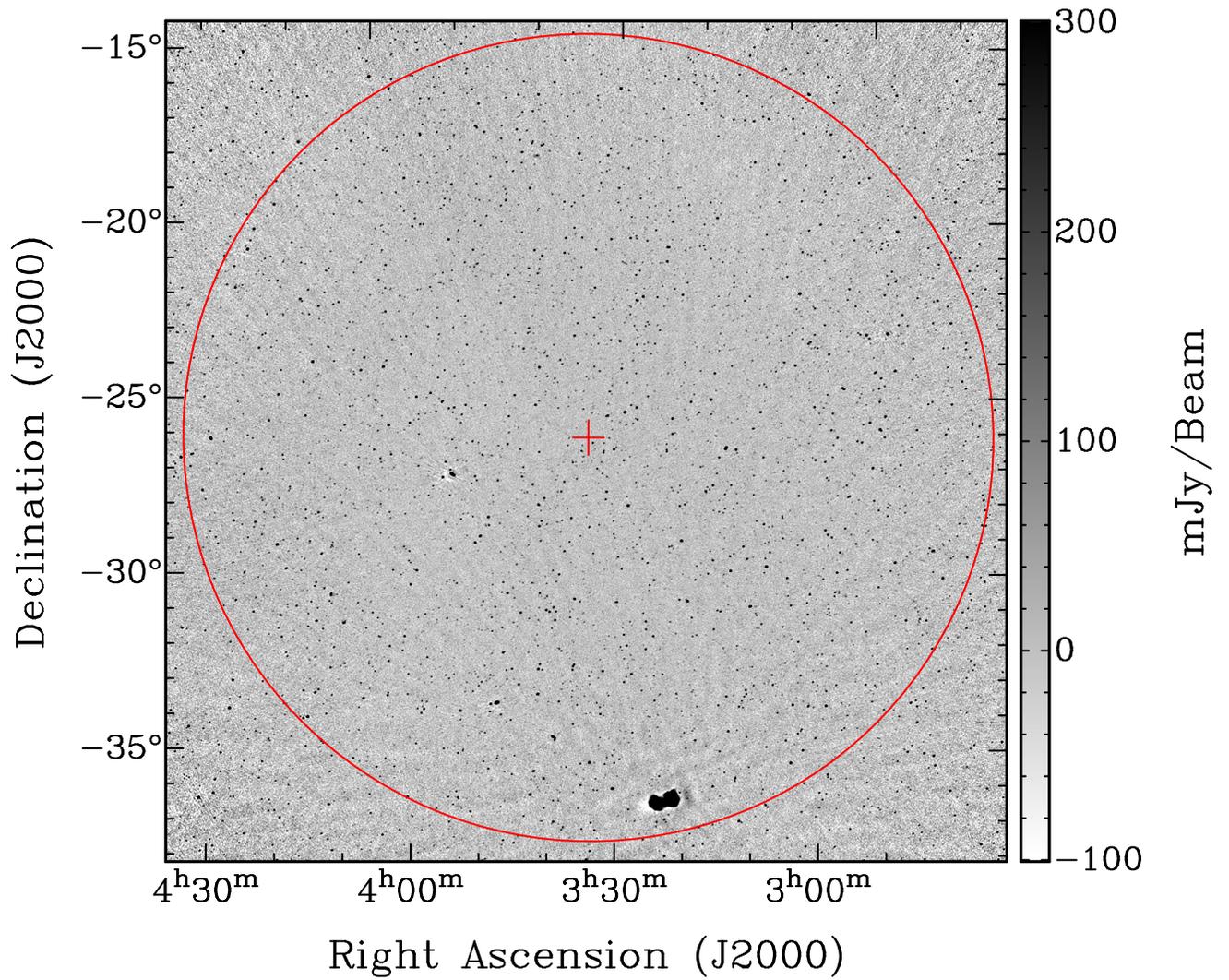}
\end{center}
\caption{GLEAM snapshot image at 139--170~MHz after primary beam correction. The image is centred close to the CDFS and Fornax A is visible in the south of the image. The red circle shows the half-power contour of the primary beam.}
\label{fig:snapshot_image}
\end{figure*}

Possible causes of the excess background noise in GLEAM images include sidelobe confusion, calibration errors, background emission from the Galactic Plane and extended sources not included in the source count model used to derive $\sigma_{\mathrm{c}}$. We analyse the noise contribution in a GLEAM snapshot image at 139--170~MHz with a beam size of 2.4~arcmin, lying close to the CDFS; the image is displayed in Fig.~\ref{fig:snapshot_image}. We then predict the visibilities for the measurement set using a realistic distribution of point sources and image the simulated $uv$ data using exactly the same parameters in \textsc{wsclean} as those used to image the real data. By comparing the $P(D)$ distributions in the real and simulated images, we show that the excess background noise is primarily caused by confusion from sidelobes of the ideal synthesized beam. Finally, we attempt to approach the theoretical noise limit using an improved deconvolution method.

%%%%%%%%%%%%%%%%%%%%%%%%%%%%%%%%%%%%%%%%%%%%%%%%%%%%%%%%%%%%%%%%%%%
\subsection{Noise properties of a real GLEAM snapshot image at 139--170~MHz}\label{Noise contribution in a real 154 MHz snapshot image}

We use the method described in Section~\ref{Estimating the theoretical noise limit} to calculate $P_{\mathrm{c}}(D) * P_{\mathrm{n}}(D)$ within the half-power contour of the primary beam. We derive $P_{\mathrm{c}}(D)$ given the beam size of 2.4~arcmin and assuming source count model A. 

\begin{figure}
\begin{center}
\includegraphics[scale=0.42,angle=270, trim=5cm 5.7cm 3cm 0cm]{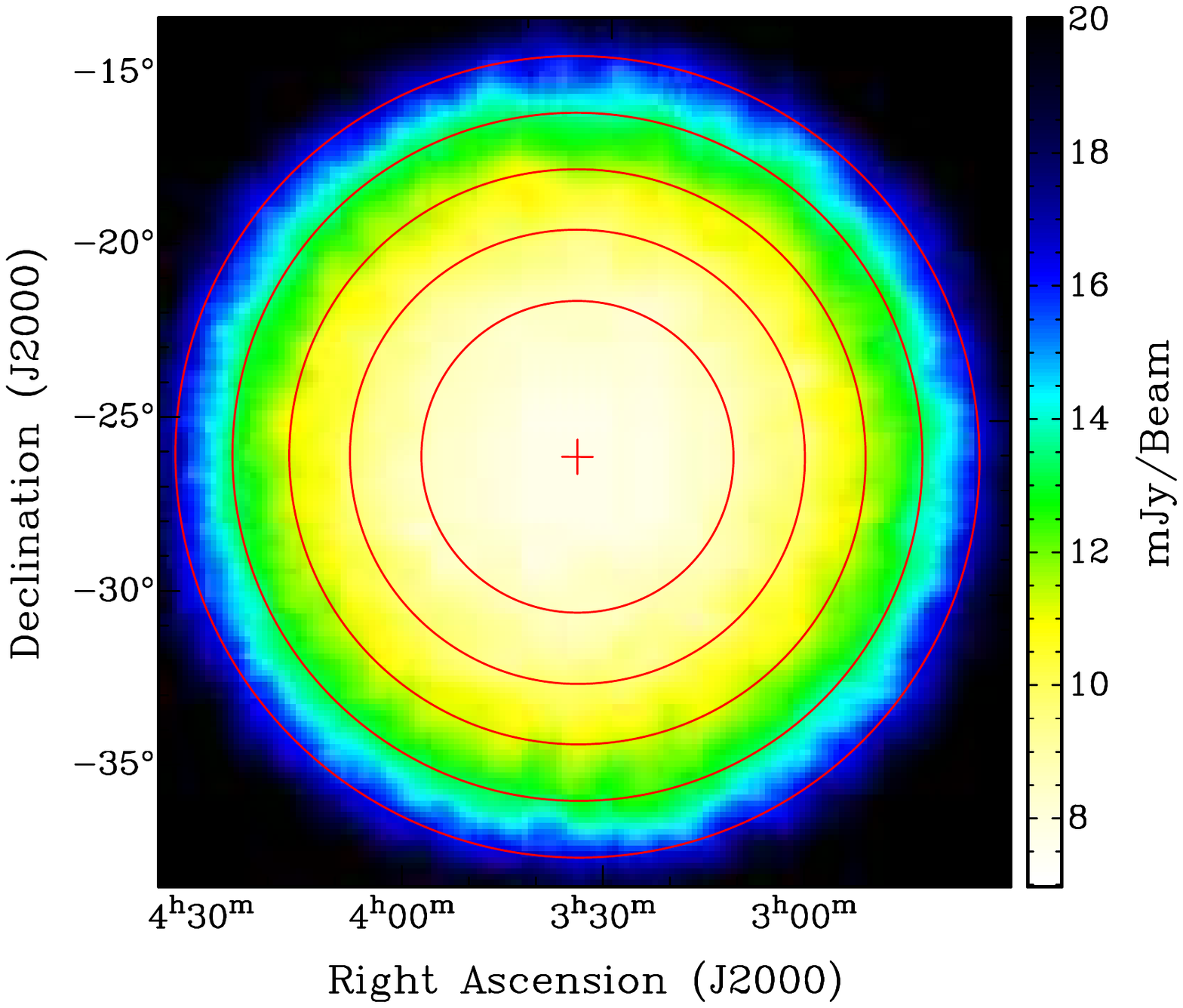}
\end{center}
\caption{Rms noise map of the Stokes $V$ snapshot image, representative of the thermal noise. The red circles show the concentric annuli into which the image was divided to calculate $P_{\mathrm{c}}(D) * P_{\mathrm{n}}(D)$.}
\label{fig:snapshot_image_rms_stokesv}
\end{figure}

Fig.~\ref{fig:snapshot_image_rms_stokesv} shows the rms noise map of the Stokes $V$ image. The thermal noise in the centre of the field is 8~mJy/beam but varies by a factor of two across the field given the primary beam response. It follows that the thermal noise distribution cannot be well approximated as a Gaussian. To address this problem, we divide the region into five concentric annuli such that the thermal noise varies by no more than 20 per cent in each annulus. The thermal noise in each annulus, $\sigma_{\mathrm{t},i}$ is taken as the mean rms noise in each annulus of the Stokes $V$ image. $P_{\mathrm{c}}(D) * P_{\mathrm{n}}(D)$ is then taken as
\begin{equation}
\frac {\sum\limits_{i=1}^{5} A_{i}~P_{\mathrm{c}}(D) * P_{\mathrm{n},i}(D) } { \sum\limits_{i=1}^{5} A_{i} }  \, ,
\end{equation}
where $P_{\mathrm{n},i}(D)$ is a Gaussian of width $\sigma_{\mathrm{t},i}$ representing the thermal noise distribution in the $i^{\mathrm{th}}$ annulus and $A_{i}$ is the area of the $i^{\mathrm{th}}$ annulus.

\begin{figure}
\begin{center}
\includegraphics[scale=0.55, trim=3.5cm 7.5cm 0cm 0cm]{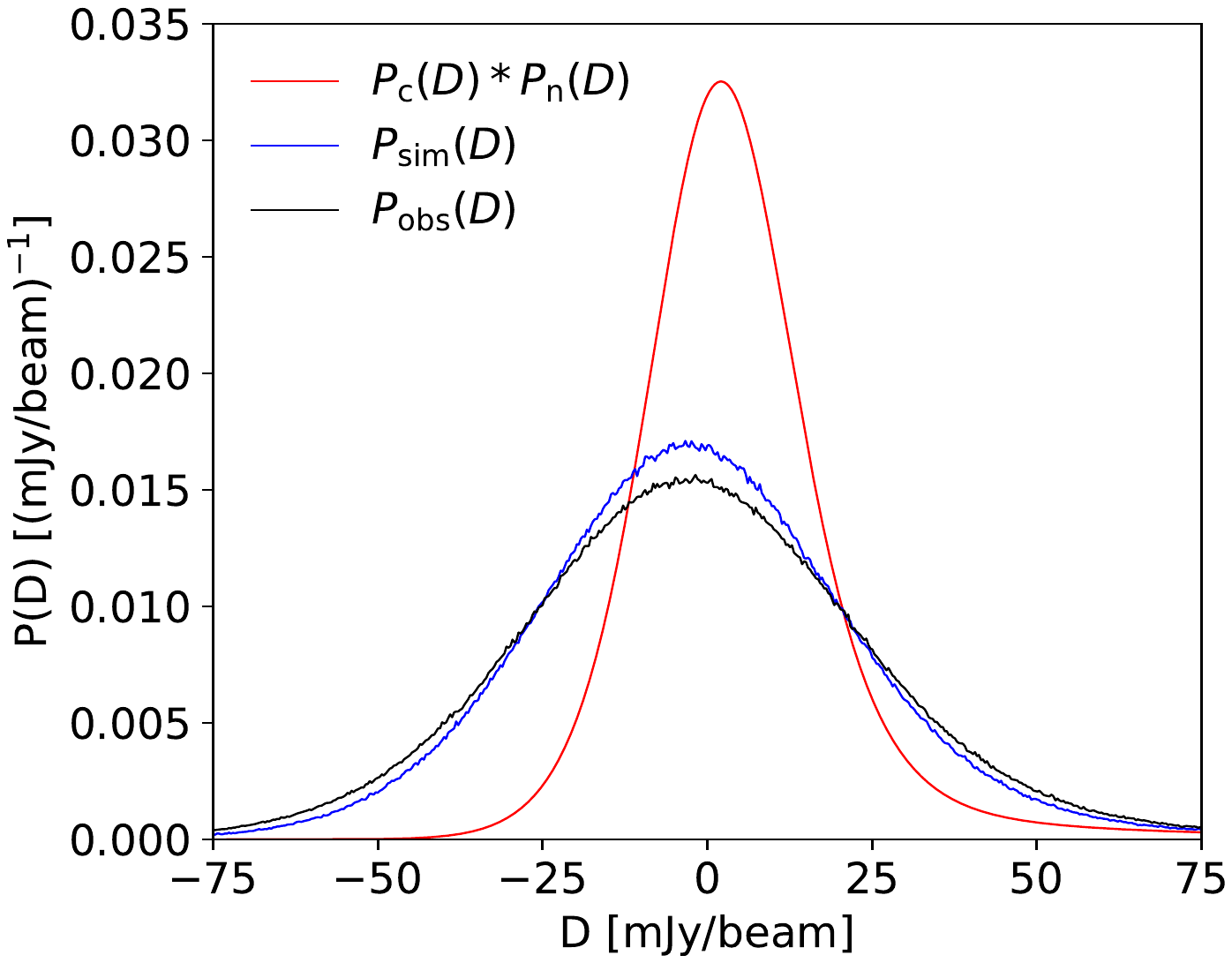}
\caption{The $P_{\mathrm{obs}}(D)$ distribution is shown in black, $P_{\mathrm{c}}(D) * P_{\mathrm{n}}(D)$ distribution in red and $P_{\mathrm{sim}}(D)$ distribution in blue.}
\label{fig:pofd_real_vs_sim}
\end{center}
\end{figure}

The observed $P(D)$ distribution within the half-power contour of the primary beam, $P_{\mathrm{obs}}(D)$, is compared with $P_{\mathrm{c}}(D) * P_{\mathrm{n}}(D)$ in Fig.~\ref{fig:pofd_real_vs_sim}. The theoretical noise limit obtained from the core width of $P_{\mathrm{c}}(D) * P_{\mathrm{n}}(D)$ is 12.7~mJy/beam. In comparison, the core width of $P_{\mathrm{obs}}(D)$, $\sigma_{\mathrm{obs}} = 26.3$~mJy/beam.

%%%%%%%%%%%%%%%%%%%%%%%%%%%%%%%%%%%%%%%%%%%%%%%%%%%%%%%%%%%%%%%%%%%
\subsection{Simulations to investigate origin of excess background noise}\label{Noise contribution in a simulated 154 MHz snapshot image}

The steps in simulating the image are as follows:
\begin{enumerate}
\item[(1)] We simulate a catalogue of point sources at 154~MHz, drawing flux densities randomly between 1~mJy and 70~Jy from source count model A. The sources lie at random positions within 40~deg from the field centre; this region is large enough to encompass the first sidelobe of the primary beam.
\item[(2)] From the simulated catalogue, we generate an image of the sky brightness distribution. Each simulated source is modelled as a $\delta$ function at the pixel closest to the source position. If more than one source is assigned to the same pixel, the flux densities of the sources are summed together. To account for the primary beam attenuation, the model image is multiplied by the primary beam response.
\item[(3)] We use the `--predict' option in \textsc{wsclean} to predict the visibilities for the measurement set from the model image.
\item[(4)] We image the simulated $uv$ data using exactly the same parameters in \textsc{wsclean} as those used to image the real data. The real image was CLEANed to 150~mJy/beam; we ensure that the simulated image is CLEANed to the same flux density threshold.
\item[(5)] We add 8~mJy/beam rms Gaussian noise to the simulated image to account for the thermal noise.
\item[(6)] We divide the simulated image by the primary beam response.
\end{enumerate}

In step 4, the simulated $uv$ data are imaged using a cell size of 32.7~arcsec, which corresponds to approximately one quarter of the synthesised beam size. Image pixelation effects coupled to the CLEAN deconvolution representation of the sky as a set of $\delta$ functions can limit the dynamic range of interferometric images \citep{cotton2008}. In order to account for this effect in the simulations, sources must be placed at various positions between cells in the simulated image. We achieve this by employing a slightly different cell size for the model image in step 2, which is used to simulate the $uv$ data.

%\textcolor{red}{Placing sources exactly on pixel centres in these simulations may yield lower noise than placing them randomly, as shown bywho investigate deconvolution artefacts arising from the use of pixelated images. For this reason, when generating the model image in step 2, we use a slightly different cell size than that used to image the simulated $uv$ data (the cell size was set to 32.7~arcsec for the simulated image and to 29.4~arcsec for the model image). This ensures that sources are located at various positions between cells in the simulated image.}

We find that the $P(D)$ distribution in the simulated image, $P_{\mathrm{sim}}(D)$, is remarkably similar to $P_{\mathrm{obs}}(D)$, as shown in Fig.~\ref{fig:pofd_real_vs_sim}. The core width of $P_{\mathrm{sim}}(D)$, $\sigma_{\mathrm{sim}} = 24.0$~mJy/beam, is only $\approx$ 9 per cent lower than $\sigma_{\mathrm{obs}}$. Since the simulated image contains no calibration artefacts, this suggests that the excess background noise in the snapshot image is primarily due to sidelobe confusion.

We repeat the simulations using source count model B but this makes negligible difference to $\sigma_{\mathrm{sim}}$. Calibration artefacts may explain the slightly higher noise level in the real image, as well as residual sidelobes from Fornax A ($S_{154~\mathrm{MHz}} = 750$~Jy; McKinley et al. 2015), which are clearly visible in the real image.

%We carried out the same analysis in the 162--170~MHz image; the results are shown in Table~\ref{tab:pofd_real_vs_sim}. The excess background noise is also due to $\sigma_{\mathrm{s}}$ at 162--170~MHz.

%down to the first negative CLEAN component. The rms noise, $\sigma$, was measured in the initial image and a new CLEAN threshold was set to 3$\sigma$. At each frequency, the image size was set to 4000 by 4000 pixels and the cell size was adjusted such that the synthesised beam was sampled by at least four pixels; the primary beam was imaged down to the $\approx 10$ per cent level at each frequency.

%%%%%%%%%%%%%%%%%%%%%%%%%%%%%%%%%%%%%%%%%%%%%%%%%%%%%%%%%%%%%%%%%%%
\subsection{Improving the deconvolution}\label{Improving the deconvolution}

The GLEAM snapshot referred to at the beginning of Section~\ref{Origin of excess background noise in GLEAM images} was imaged using \textsc{wsclean} v1.10. The pixel size was set to $32.7 \times 32.7~\mathrm{arcsec}^{2}$ and the image size to $4000 \times 4000$ pixels, such that the image encompasses the $\approx 10$ per cent level of the primary beam. The snapshot was imaged down to the first negative CLEAN component. The rms noise of this initial image, $\sigma = 50$~mJy/beam, was measured and the snapshot was re-imaged down to a CLEAN threshold of $3\sigma$ (150~mJy/beam). In practice, this CLEANing strategy generally leaves significant residual emission undeconvolved.

We re-image the snapshot using \textsc{wsclean} v2.5, which is more efficient for large images thanks to the implementation of the Clark CLEAN algorithm \citep{clark1980}. In minor CLEAN cycles, CLEAN components are subtracted from the image using only the central portion of the PSF and only the largest residuals are searched. This is sufficient to find the CLEAN components providing that the synthesised beam is well behaved; the accuracy of the subtraction is improved during major CLEAN cycles where the FT of the CLEAN components is subtracted from the residual visibility data.

Using \textsc{wsclean} v2.5, we CLEAN the entire image to 3$\sigma$, construct a mask from the identified CLEAN components and continue CLEANing with the mask to $1\sigma$. This is conducted in an automated fashion using the `auto-mask' and `auto-threshold' parameters. It is not necessary to provide \textsc{wsclean} with an estimate of $\sigma$ as the algorithm automatically calculates the standard deviation of the residual image before the start of every major CLEAN cycle, which it then uses to set the CLEAN threshold. This is desirable since, in practice, the noise can drop considerably after the first few major CLEAN cycles as the image quality improves. The use of a mask permits CLEANing down to the noise level.

\begin{figure}
\begin{center}
\includegraphics[scale=0.45, trim=1cm 6cm 0cm 0cm]{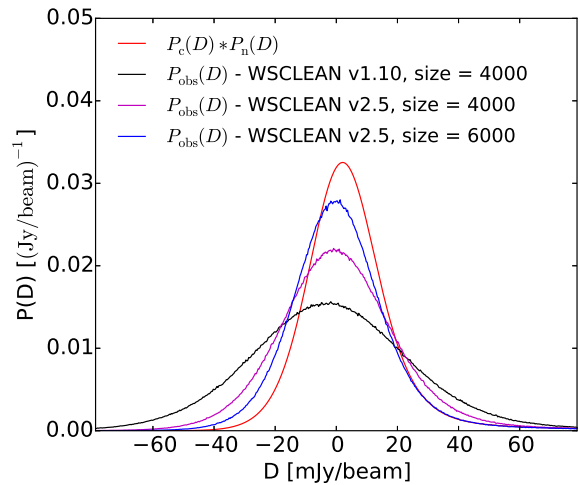}
\caption{Observed $P(D)$ distributions obtained using different versions of \textsc{wsclean} and image sizes. The theoretical noise limit is shown in red.}
\label{fig:deeper_clean}
\end{center}
\end{figure}

The total number of CLEAN iterations using \textsc{wsclean} v2.5 is $\approx 190,000$ while it is only $\approx 25,000$ using \textsc{wsclean} v1.10. Despite the much larger number of CLEAN iterations, the processing time for \textsc{wsclean} v2.5 is $\approx$ 4 times shorter. The $P(D)$ distributions obtained using the two versions of \textsc{wsclean} are compared with the theoretical noise limit in Fig.~\ref{fig:deeper_clean}. There is a $\approx 29$ per cent reduction in $\sigma_{\mathrm{obs}}$ using \textsc{wsclean} v2.5.

We investigate whether the noise can be reduced further by increasing the size of the region being CLEANed. We re-run \textsc{wsclean} v2.5 increasing the image size from $4000 \times 4000$ to $6000 \times 6000$ pixels. The imaged field-of-view now encompasses the first null of the primary beam. The resulting $P(D)$ distribution is displayed in Fig.~\ref{fig:deeper_clean}. There is a further $\approx 21$ per cent reduction in $\sigma_{\mathrm{obs}}$, which is now only $\approx 15$ per cent above $\sigma_{\mathrm{lim}}$.

The results of this analysis are summarised in Table~\ref{tab:deeper_clean}. We conclude that both the limited CLEANing depth and far-field sources that have not been deconvolved contribute significantly to the sidelobe confusion in GLEAM. The Clark optimisation is highly effective for large MWA images, permitting deeper CLEANing. We recommend adopting this technique in the future to ensure full exploitation of MWA survey images with the auto-masking and deeper thresholding.

%given the wall clock time limit of 12 hours on the supercomputer.
 
% Tables from:
% ../sim_1069428552/final_0003/refine/pofd/noise_stats.txt
% ../sim_1069428552/final_MFS/refine/pofd/noise_stats.txt
% ADD PROCESSING TIME
\begin{table*}
\centering
\caption{Key parameters recorded for three different runs of \textsc{wsclean} on a 154~MHz snapshot image (see text for details). The theoretical noise limit is 12.7~mJy/beam.}
\label{tab:deeper_clean}
\begin{tabular}{@{} c c c c c} 
\hline
\textsc{wsclean}
& Image size
& Number of
& $\sigma_{\mathrm{obs}}$
&Processing time\\
version
& (pixels)
& CLEAN iterations
& (mJy/beam)
& (hours) \\
\hline
1.10 & 4000 & $\approx 25,000$ & 26.3 & 5.0 \\
2.5 & 4000 & $\approx 193,000$ & 18.6 & 1.2 \\
2.5 & 6000 & $\approx 500,000$ & 14.7 & 10.1 \\
\hline
\end{tabular}
\end{table*}

%%%%%%%%%%%%%%%%%%%%%%%%%%%%%%%%%%%%%%%%%%%%%%%%%%%%%%%%%%%%%%%%%%%
\section{Prospects for MWA phase 2}\label{Prospects for MWA phase 2}

Since the GLEAM survey observations were carried out, the MWA has been upgraded with the addition of a further 128 tiles, 56 of which lie on baselines up to $\approx 6$~km, roughly improving the array resolution by a factor of two \citep{wayth2018}. The correlator capacity was not increased in Phase 2 of the MWA, so it is still only possible to correlate 128 tiles. In this section, we give an overview of how we expect $\sigma_{\mathrm{c}}$ and the rms sidelobe confusion noise, $\sigma_{\mathrm{s}}$, to change for MWA Phase 2 observations.

\begin{figure}
\begin{center}
\includegraphics[scale=0.35, trim=3cm 2cm 0cm 0cm]{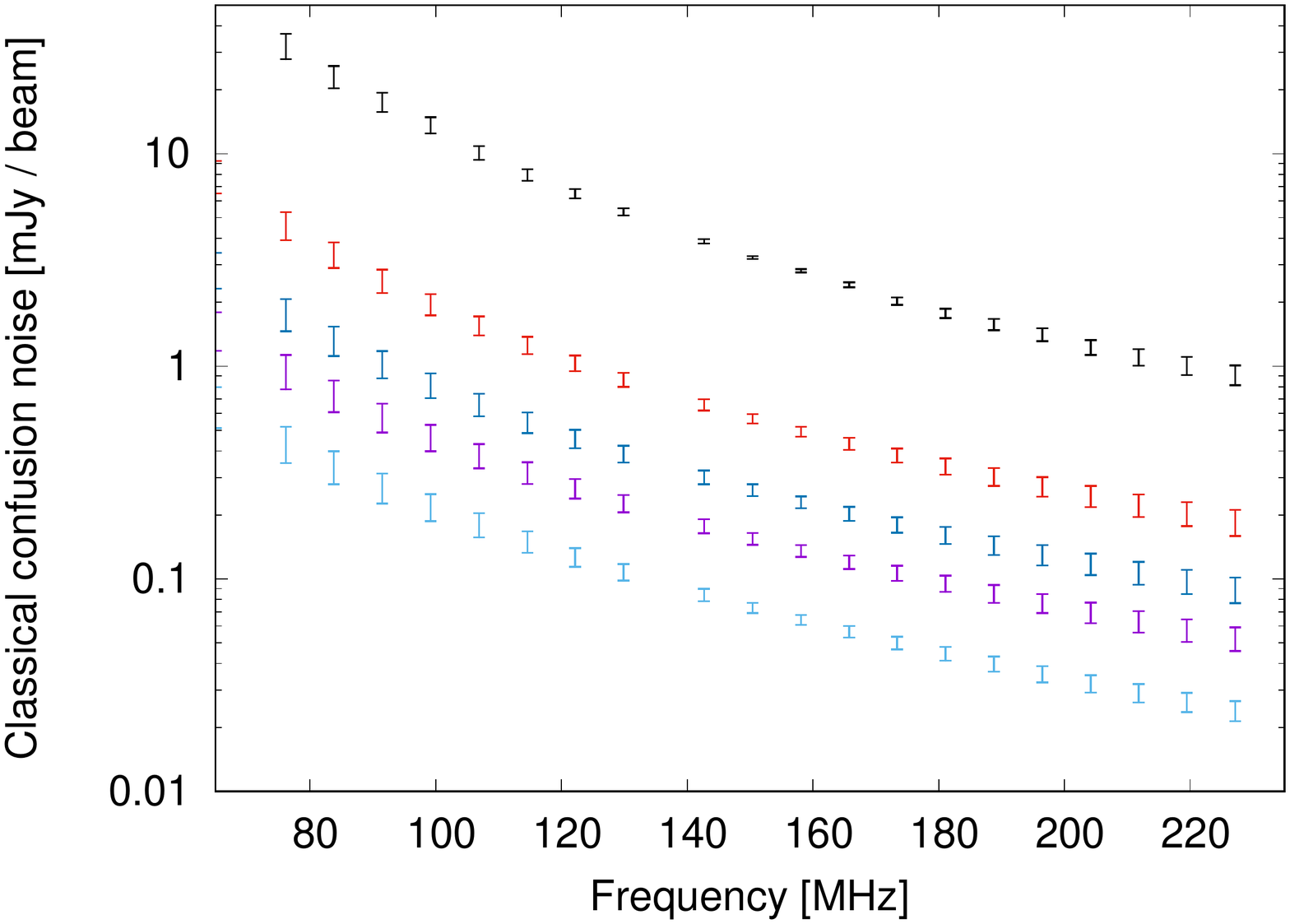}
\includegraphics[scale=0.35, trim=3cm 2cm 0cm 0cm]{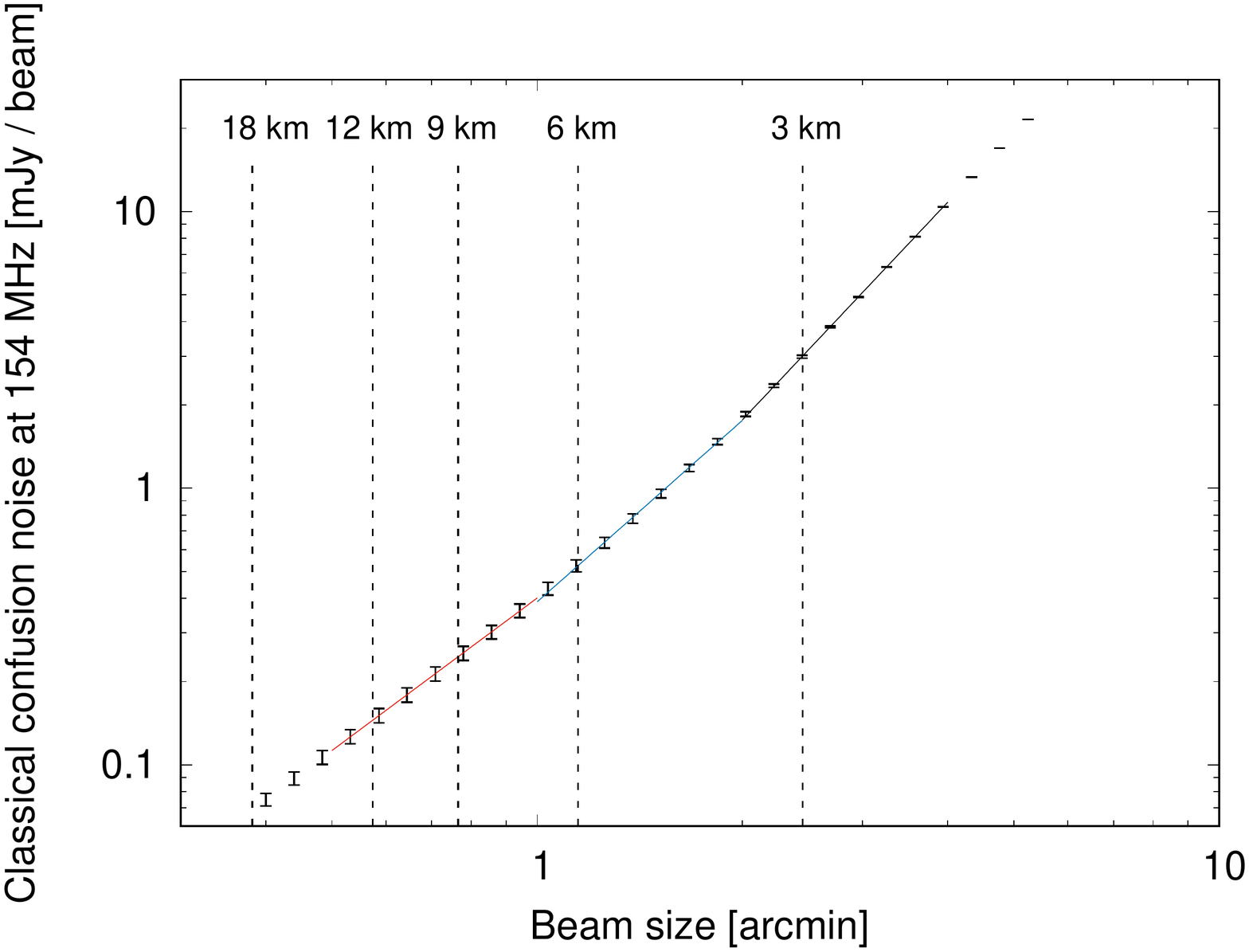}
\caption{Top: classical confusion noise as a function of frequency for MWA Phase 1 (black), MWA Phase 2 (red) and larger, hypothetical arrays with maximum baselines of 9~km (blue), 12~km (purple) and 18~km (turquoise). Bottom: classical confusion noise at 154~MHz as a function of beam size. The diagonal lines show power-law fits to the data points in three different $\theta$ ranges. Dashed lines indicate the beam sizes at 154~MHz for the different arrays.}
\label{fig:classical_conf_future}
\end{center}
\end{figure}

We use the \textsc{miriad} \citep{sault1995} task \textsc{uvgen} to simulate an image of the MWA Phase 2 PSF for a 2~min snapshot with a central frequency of 154~MHz and a bandwidth of 30.72~MHz, using a uniform weighting scheme. We fit a Gaussian to the main lobe of the synthesised beam; the geometric average of the major and minor axes of the fitted Gaussian is 1.15~arcmin. We use the method described in Section~\ref{Estimating the theoretical noise limit} to derive $\sigma_{\mathrm{c}}$ as a function of frequency for MWA Phase 2, setting the beam size $\theta_{\mathrm{Phase~2}} = 1.15~\mathrm{arcmin}~(\nu/154~\mathrm{MHz})^{-1}$. We find that $\sigma_{\mathrm{c,~Phase~1}}/\sigma_{\mathrm{c,~Phase~2}}$ varies from $\approx 5$ at the high end of the band to $\approx 7$ at the low end of the band, as shown in the top panel of Fig.~\ref{fig:classical_conf_future}. The top panel of Fig.~\ref{fig:classical_conf_future} also includes estimates of $\sigma_{\mathrm{c}}$ for larger, hypothetical arrays with maximum baselines of 9, 12 and 18~km, where we set the beam size to $\frac{2}{3} \theta_{\mathrm{Phase~2}}$, $\frac{1}{2} \theta_{\mathrm{Phase~2}}$ and $\frac{1}{3} \theta_{\mathrm{Phase~2}}$, respectively.

The classical confusion noise at 154~MHz as a function of beam size is also displayed in the bottom panel of Fig.~\ref{fig:classical_conf_future}. We fit the function $\sigma_{\mathrm{c}} = a \theta^{b}$ in three different $\theta$ ranges and find that $b$ drops with decreasing $\theta$, with $b = 2.61$ for $\theta = 2.0-4.0$~arcmin, $b = 2.18$ for $\theta = 1.0-2.0$~arcmin and $b = 1.83$ for $\theta = 0.5-1.0$~arcmin. \cite{condon1974} showed that for a power-law differential source count $n(S) = k S^{-\gamma}$, $\sigma_{\mathrm{c}} \propto \theta^\frac{2}{\gamma-1}$. The flattening of the 154~MHz Euclidean normalised differential counts below $\approx 10$~mJy (corresponding to an increase in $\gamma$), can therefore explain the drop in $b$ with decreasing $\theta$.

\cite{bowman2009} derive an expression for the variance in the intensity of a dirty sky map assuming that the primary and synthesised beams are described by top-hat functions, such that the response is defined to be one within a region of diameter $\Theta_{\mathrm{P}}$ in the case of the primary beam and within a region of diameter $\Theta_{\mathrm{B}}$ in the case of the synthesised beam. Outside this region, the response is taken to be zero for the primary beam and a constant value of $B_{\mathrm{rms}} \ll 1$ for the synthesised beam, representing the standard deviation of the synthesised beam sidelobes. With these simplifications,
\begin{equation}
\sigma_{\mathrm{s}} \approx \sigma_{\mathrm{c}} B_{\mathrm{rms}} \sqrt \frac{\Omega_{\mathrm{P}}}{\Omega_{\mathrm{B}}} \, ,
\end{equation}
where $\Omega_{\mathrm{B}} \approx \Theta_{\mathrm{B}}^{2}$ is the solid angle of the synthesised beam and $\Omega_{\mathrm{P}} \approx \Theta_{\mathrm{P}}^{2}$ is the solid angle of the primary beam.

\begin{figure}
\begin{center}
\includegraphics[scale=0.35, trim=3cm 2cm 0cm 0cm]{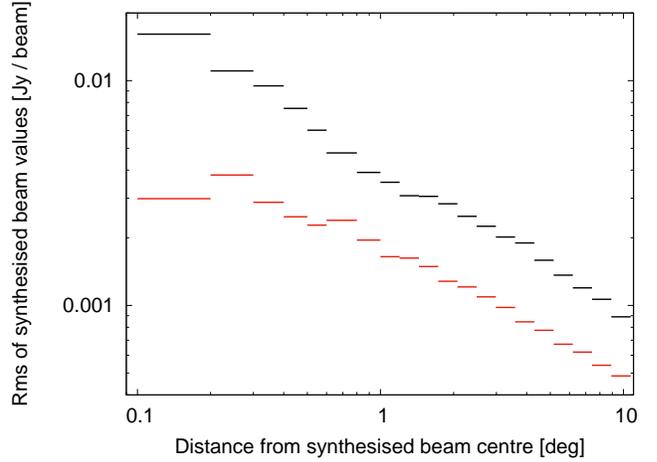}
\caption{Comparison of the MWA phase 1 (black) and 2 (red) synthesised beams for a 2~min snapshot with a central frequency of 154~MHz and bandwidth of 30.72~MHz, using a uniform weighting scheme. The standard deviation of the pixel values in the synthesised beam is plotted as a function of distance from the pointing centre. The standard deviation is calculated in a thin annulus at the given radius.}
\label{fig:rms_vs_distance}
\end{center}
\end{figure}

For the MWA, $B_{\mathrm{rms}}$ varies strongly with distance, $d$, from the beam centre. We measure $B_{\mathrm{rms}}$ in MWA Phase 1 and 2 PSF images as a function of $d$ (see Fig.~\ref{fig:rms_vs_distance}). Since $B_{\mathrm{rms,~Phase~1}} \gtrapprox 2~B_{\mathrm{rms,~Phase~2}}$, $\Omega_{\mathrm{P,~Phase~1}} = \Omega_{\mathrm{P,~Phase~2}}$ and $\Omega_{\mathrm{B,~Phase~1}} \approx 4~\Omega_{\mathrm{B,~Phase~2}}$,
\begin{equation}
\frac{\sigma_{\mathrm{s,~Phase~1}}}{\sigma_{\mathrm{s,~Phase~2}}} \gtrapprox \frac{\sigma_{\mathrm{c,~Phase~1}}}{\sigma_{\mathrm{c,~Phase~2}}} \, .
\end{equation}

We therefore expect that $\sigma_{\mathrm{s,~Phase~1}}/\sigma_{\mathrm{s,~Phase~2}} \gtrapprox 5$ across the MWA frequency range, assuming that the MWA Phase 1 and 2 images are CLEANed to the same flux density threshold. We must also consider that MWA Phase 2 images will take longer to image because of the increased resolution. The calibration of MWA Phase 2 data will be more challenging and, depending on the ionospheric conditions, direction-dependent calibration techniques will probably be required to reach the theoretical noise limit (see e.g. Offringa et al. 2016, Rioja, Dodson \& Franzen, submitted).

%%%%%%%%%%%%%%%%%%%%%%%%%%%%%%%%%%%%%%%%%%%%%%%%%%%%%%%%%%%%%%%%%%%
%\newpage
\section{Summary and future work}\label{Summary and future work}

GLEAM is a contiguous 72--231~MHz survey of the entire sky south of declination~$+ 30^{\circ}$ and has the widest fractional bandwidth and highest surface brightness sensitivity among low radio frequency surveys. We have determined the GLEAM source counts at 200, 154, 118 and 88~MHz to a flux density limit of 50, 80, 120 and 290~mJy respectively, to high precision. The 200~MHz counts are based on the GLEAM extragalactic catalogue by \cite{hurleywalker2017}. From the three lowest 30.72~MHz sub-band images of GLEAM, we have constructed additional, statistically complete source samples at 154, 118 and 88~MHz to measure the counts at these frequencies.

The counts at 154 and 88~MHz are overall in good agreement with other counts in the literature at a similar frequency. The 151~MHz SKADS model significantly underpredicts the 154~MHz GLEAM counts at $S \gtrsim 50$~mJy. The cause of the discrepancy is unclear. The model is based on the 151~MHz luminosity function of high-luminosity radio galaxies by \cite{willott2001}, which in turn was determined using measurements of the local radio luminosity function (LRLF) for AGN. Since no measurements of the LRLF for AGN were available at 151~MHz, \citeauthor{willott2001} used the LRLF for AGN at 1.4~GHz by \cite{cotton1998} and made a simple shift in radio power assuming $\alpha = -0.8$. They also chose to fit a Schechter luminosity function, whose exponential high-luminosity cutoff is likely too sharp to describe radio galaxies. We find that the model is statistically in much better agreement with the data after multiplying the simulated flux densities by 1.2.

%\citeauthor{willott2001} extrapolated the 1.4~GHz local RLF for AGN by \cite{cotton1998} to 151~MHz. 
%may prevent the model from accurately reproducing the source counts at the bright end.
%The reliability of this extrapolation is questionable given the large frequency difference. 

%It may be due to the limited volume covered by the model, biasing against rare, bright sources. 
%Cotton \& Condon (1998)

At $S_{154~\mathrm{MHz}} > 0.5$~Jy, there is no discernible change in the shape of the counts at the four frequencies: a spectral index scaling of $\approx -0.8$ provides a good match between the counts. The spectra of individual sources show, on average, a slight but significant flattening of $\delta \alpha_{76}^{227} \approx 0.1$ between 0.5 and 0.1~Jy.

We may have expected to see a change in the source count shape with frequency due to spectral curvature of generations of sources at different redshifts. The fact that GLEAM is overwhelmingly dominated by sources with steep, power-law spectra indicates that there is no simple way of tracing ageing or evolution of the bright source population from this set of frequencies.

The low-frequency emission from star-forming galaxies remains largely unstudied. Detailed measurements of their spectra are important for understanding the physical processes which contribute to the radio emission from star formation. They can also be used to construct more accurate low frequency source counts, which will be invaluable for planning deep low-frequency surveys with future facilities. \cite{galvin2018} measured the radio spectra of 19 luminous infrared galaxies (LIRGs) at $0.067 < z < 0.227$ using GLEAM and Australia Telescope Compact Array (ATCA) follow-up observations at 2.1--45~GHz. They found that many of the sources exhibit low-frequency turnovers in their spectra which can be attributed, in large part, to free-free absorption. Deep LOFAR observations in small-area fields are also probing the low frequency behaviour of star-forming galaxies. The LoTSS is expected to detect hundreds of thousands of star-forming galaxies, primarily at lower redshifts but extending out to $z \geq 1$.

%\textcolor{red}{\cite{kapinska2017} studied the spectral energy distribution of NGC~253 between 76~MHz and 11~GHz. The spectrum was best described as the sum of a central starburst, modelled as an internally free-free absorbed synchrotron plasma, and an extended synchrotron component flattening at low radio frequencies.} 

Although GLEAM is overwhelmingly dominated by radio-loud AGN, the SKADS model predicts that GLEAM contains $375 \pm 80$ local ($z < 0.1$) star-forming galaxies with $S_{200~\mathrm{MHz}} > 50$~mJy in region B, covering $\approx 6500~\mathrm{deg}^{2}$. In a future paper, we will cross-match the GLEAM catalogue with nearby optical samples to determine the LRLF for both AGN and star-forming galaxies at 154~MHz. We will correlate the local radio sample with higher frequency surveys including NVSS and SUMSS to characterise the typical spectra of these two populations. We also plan to investigate changes in the spectral behaviour of AGN with respect to radio morphology and luminosity.

Using deep 150~MHz LOFAR counts by \cite{williams2016} and the SKADS model, we have conducted a $P(D)$ analysis to derive the classical confusion noise in GLEAM images. While the images are limited by classical confusion below $\approx 100$~MHz, the rms noise is a factor of $\approx 2$ higher than the theoretical noise limit, accounting for both the thermal noise and classical confusion, at higher frequencies. By analysing a synthetic snapshot image containing a realistic distribution of point sources, we have demonstrated that the excess background noise is primarily due to confusion from sidelobes of the ideal synthesized beam. We have shown that we can approach the theoretical noise limit using the Clark CLEAN algorithm implemented in \textsc{wsclean}, along with deeper deconvolution and larger image size to encompass the first null of the primary beam.

For the MWA Phase 2 array with the angular resolution improved by a factor of two, we anticipate that both the classical and sidelobe confusion noise will drop by a factor of $\approx 5$ at the high end of the band. Deep pointed observations of the Galaxy and Mass Assembly \citep[GAMA;][]{driver2009} 23 field, centred at Dec~$- 32.5^{\circ}$, have been made with MWA Phase 2 (Seymour et al., in preparation) at 72--231~MHz with the goal of producing a radio luminosity function and investigating its dependence on MWA in-band spectral index. This work will demonstrate the `deep' imaging quality which MWA Phase 2 can provide and will include an investigation of the factors which affect the noise.

%%%%%%%%%%%%%%%%%%%%%%%%%%%%%%%%%%%%%%%%%%%%%%%%%%%%%%%%%%%%%%%%%%%
\begin{acknowledgements}
This scientific work makes use of the Murchison Radio-astronomy Observatory, operated by CSIRO. We acknowledge the Wajarri Yamatji people as the traditional owners of the Observatory site. Support for the operation of the MWA is provided by the Australian Government (NCRIS), under a contract to Curtin University administered by Astronomy Australia Limited. We thank the anonymous referee for helpful comments, which have substantially improved this paper. We acknowledge the Pawsey Supercomputing Centre which is supported by the Western Australian and Australian Governments. CAJ thanks the Department of Science, Office of Premier \& Cabinet, WA for their support through the Western Australian Fellowship Program.
\end{acknowledgements}

%%%%%%%%%%%%%%%%%%%%%%%%%%%%%%%%%%%%%%%%%%%%%%%%%%%%%%%%%%%%%%%%%%%
\begin{appendix}
\section{Source count data}\label{Source count data}

The 200, 154, 118 and 88~MHz source count data presented in this paper are provided in Table~\ref{tab:gleam_source_counts}.

\begin{table*}
\small
\centering
\caption{Euclidean normalised differential source counts for GLEAM at 200, 154, 118 and 88~MHz. The bin centre corresponds to the mean flux density of all sources in the bin. The quoted counts are corrected for incompleteness, Eddington bias and source blending as described in the text; the correction factor for each bin is provided for reference.}
\label{tab:gleam_source_counts}
\begin{tabular}{@{} c c c c c c c c} 
\hline
Frequency
&\multicolumn{1}{c}{Bin start}
&\multicolumn{1}{c}{Bin end}
&\multicolumn{1}{c}{Bin centre}
&\multicolumn{1}{c}{Raw number}
&Euclidean normalised
&\multicolumn{1}{c}{Correction}
&Region\\
(MHz)
&\multicolumn{1}{c}{$S$ (Jy)}
&\multicolumn{1}{c}{$S$ (Jy)}
&\multicolumn{1}{c}{$S$ (Jy)}
&\multicolumn{1}{c}{of sources, $N$}
&\multicolumn{1}{c}{counts ($\mathrm{Jy}^{3/2} \mathrm{sr}^{-1}$)}
&\multicolumn{1}{c}{factor}
& \\
\hline
200 & 0.044 & 0.055 & 0.0493 & 13864 & $ 378 \pm 8 $ & $ 1.10 \pm 0.02 $ & B \\
& 0.055 & 0.069 & 0.0616 & 12919 & $ 465 \pm 10 $ & $ 1.06 \pm 0.02 $ & B \\
& 0.069 & 0.086 & 0.0771 & 11339 & $ 575 \pm 12 $ & $ 1.04 \pm 0.02 $ & B \\
& 0.086 & 0.107 & 0.0959 & 10210 & $ 711 \pm 16 $ & $ 1.02 \pm 0.02 $ & B \\
& 0.107 & 0.134 & 0.1199 & 28801 & $ 802 \pm 15 $ & $ 1.14 \pm 0.02 $ & A \\
& 0.134 & 0.168 & 0.1501 & 26880 & $ 965 \pm 19 $ & $ 1.06 \pm 0.02 $ & A \\
& 0.168 & 0.210 & 0.1879 & 23025 & $ 1137 \pm 23 $ & $ 1.03 \pm 0.02 $ & A \\
& 0.210 & 0.262 & 0.2343 & 19690 & $ 1342 \pm 28 $ & $ 1.01 \pm 0.02 $ & A \\
& 0.262 & 0.328 & 0.2928 & 16810 & $ 1541 \pm 14 $ & $ 0.99 \pm 0.01 $ & A \\
& 0.328 & 0.410 & 0.3664 & 13791 & $ 1774 \pm 18 $ & $ 0.98 \pm 0.01 $ & A \\
& 0.410 & 0.512 & 0.4571 & 11041 & $ 1976 \pm 21 $ & $ 0.98 \pm 0.01 $ & A \\
& 0.512 & 0.640 & 0.5712 & 8721 & $ 2178 \pm 27 $ & $ 0.98 \pm 0.01 $ & A \\
& 0.640 & 0.800 & 0.7120 & 6786 & $ 2353 \pm 33 $ & $ 0.98 \pm 0.01 $ & A \\
& 0.800 & 1.000 & 0.8912 & 5190 & $ 2494 \pm 39 $ & $ 0.97 \pm 0.01 $ & A \\
& 1.000 & 1.250 & 1.1160 & 4009 & $ 2736 \pm 46 $ & $ 0.98 \pm 0.01 $ & A \\
& 1.250 & 1.560 & 1.3909 & 2971 & $ 2812 \pm 55 $ & $ 0.97 \pm 0.01 $ & A \\
& 1.560 & 1.950 & 1.7417 & 2094 & $ 2796 \pm 65 $ & $ 0.98 \pm 0.01 $ & A \\
& 1.950 & 2.440 & 2.1702 & 1520 & $ 2825 \pm 76 $ & $ 0.99 \pm 0.01 $ & A \\
& 2.440 & 3.050 & 2.7023 & 1124 & $ 2848 \pm 89 $ & $ 0.97 \pm 0.01 $ & A \\
& 3.050 & 3.820 & 3.3927 & 701 & $ 2561 \pm 109 $ & $ 1.00 \pm 0.02 $ & A \\
& 3.820 & 4.770 & 4.2372 & 461 & $ 2361 \pm 116 $ & $ 1.00 \pm 0.02 $ & A \\
& 4.770 & 5.960 & 5.2773 & 333 & $ 2314 \pm 129 $ & $ 0.98 \pm 0.01 $ & A \\
& 5.960 & 7.450 & 6.5870 & 232 & $ 2269 \pm 152 $ & $ 0.99 \pm 0.01 $ & A \\
& 7.450 & 9.310 & 8.2935 & 152 & $ 2165 \pm 181 $ & $ 1.01 \pm 0.02 $ & A \\
& 9.310 & 11.600 & 10.3344 & 101 & $ 2001 \pm 199 $ & - & A \\
& 11.600 & 14.600 & 13.0278 & 61 & $ 1646 \pm 211 $ & - & A \\
& 14.600 & 18.200 & 16.0634 & 55 & $ 2088 \pm 282 $ & - & A \\
& 18.200 & 22.700 & 20.4399 & 25 & $ 1387 \pm 277 $ & - & A \\
& 22.700 & 28.400 & 24.6357 & 12 & $ 838 \pm 242 $ & - & A \\
& 28.400 & 56.800 & 41.3552 & 20 & $ 1024 \pm 229 $ & - & A \\
& 56.800 & 113.700 & 75.7906 & 3 & $ 348^{+341}_{-189} $ & - & A \\
\hline
154 & 0.069 & 0.086 & 0.0772 & 11193 & $ 601 \pm 12 $ & $ 1.09 \pm 0.02 $ & B \\
& 0.086 & 0.107 & 0.0959 & 10216 & $ 760 \pm 16 $ & $ 1.09 \pm 0.02 $ & B \\
& 0.107 & 0.134 & 0.1198 & 9409 & $ 909 \pm 20 $ & $ 1.04 \pm 0.02 $ & B \\
& 0.134 & 0.168 & 0.1502 & 27363 & $ 1071 \pm 20 $ & $ 1.15 \pm 0.02 $ & A \\
& 0.168 & 0.210 & 0.1879 & 25309 & $ 1285 \pm 26 $ & $ 1.05 \pm 0.02 $ & A \\
& 0.210 & 0.262 & 0.2346 & 22288 & $ 1529 \pm 32 $ & $ 1.01 \pm 0.02 $ & A \\
& 0.262 & 0.328 & 0.2930 & 19560 & $ 1800 \pm 16 $ & $ 0.99 \pm 0.01 $ & A \\
& 0.328 & 0.410 & 0.3664 & 16366 & $ 2099 \pm 20 $ & $ 0.98 \pm 0.01 $ & A \\
& 0.410 & 0.512 & 0.4577 & 13462 & $ 2420 \pm 27 $ & $ 0.98 \pm 0.01 $ & A \\
& 0.512 & 0.640 & 0.5713 & 10764 & $ 2674 \pm 31 $ & $ 0.97 \pm 0.01 $ & A \\
& 0.640 & 0.800 & 0.7136 & 8553 & $ 2933 \pm 37 $ & $ 0.96 \pm 0.01 $ & A \\
& 0.800 & 1.000 & 0.8906 & 6629 & $ 3194 \pm 44 $ & $ 0.97 \pm 0.01 $ & A \\
& 1.000 & 1.250 & 1.1148 & 5097 & $ 3465 \pm 53 $ & $ 0.98 \pm 0.01 $ & A \\
& 1.250 & 1.560 & 1.3935 & 3806 & $ 3578 \pm 62 $ & $ 0.96 \pm 0.01 $ & A \\
& 1.560 & 1.950 & 1.7365 & 2849 & $ 3783 \pm 76 $ & $ 0.99 \pm 0.01 $ & A \\
& 1.950 & 2.440 & 2.1738 & 2022 & $ 3690 \pm 88 $ & $ 0.97 \pm 0.01 $ & A \\
& 2.440 & 3.050 & 2.7123 & 1501 & $ 3869 \pm 106 $ & $ 0.98 \pm 0.01 $ & A \\
& 3.050 & 3.820 & 3.3869 & 1106 & $ 3932 \pm 124 $ & $ 0.98 \pm 0.01 $ & A \\
& 3.820 & 4.770 & 4.2445 & 651 & $ 3311 \pm 138 $ & $ 0.98 \pm 0.01 $ & A \\
& 4.770 & 5.960 & 5.3171 & 457 & $ 3263 \pm 158 $ & $ 0.99 \pm 0.01 $ & A \\
& 5.960 & 7.450 & 6.6146 & 316 & $ 3097 \pm 178 $ & $ 0.98 \pm 0.01 $ & A \\
\end{tabular}
\end{table*}

\begin{table*}
\small
%\contcaption{}
\centering
\begin{tabular}{@{} c c c c c c c c} 
\hline
Frequency
&\multicolumn{1}{c}{Bin start}
&\multicolumn{1}{c}{Bin end}
&\multicolumn{1}{c}{Bin centre}
&\multicolumn{1}{c}{Raw number}
&Euclidean normalised
&\multicolumn{1}{c}{Correction}
&Region\\
(MHz)
&\multicolumn{1}{c}{$S$ (Jy)}
&\multicolumn{1}{c}{$S$ (Jy)}
&\multicolumn{1}{c}{$S$ (Jy)}
&\multicolumn{1}{c}{of sources, $N$}
&\multicolumn{1}{c}{counts ($\mathrm{Jy}^{3/2} \mathrm{sr}^{-1}$)}
&\multicolumn{1}{c}{factor}
& \\
\hline
154 & 7.450 & 9.310 & 8.2454 & 223 & $ 3053 \pm 209 $ & $ 0.99 \pm 0.01 $ & A \\
& 9.310 & 11.600 & 10.3656 & 133 & $ 2656 \pm 230 $ & - & A \\
& 11.600 & 14.600 & 12.8897 & 104 & $ 2733 \pm 268 $ & - & A \\
& 14.600 & 18.200 & 16.4515 & 56 & $ 2257 \pm 302 $ & - & A \\
& 18.200 & 22.700 & 19.7413 & 49 & $ 2492 \pm 356 $ & - & A \\
& 22.700 & 28.400 & 25.2264 & 30 & $ 2223 \pm 406 $ & - & A \\
& 28.400 & 56.800 & 40.6881 & 24 & $ 1180 \pm 241 $ & - & A \\
& 56.800 & 113.700 & 75.3991 & 7 & $ 803^{+434}_{-296} $ & - & A \\
\hline
118 & 0.107 & 0.134 & 0.1202 & 9139 & $ 994 \pm 20 $ & $ 1.16 \pm 0.02 $ & B \\
& 0.134 & 0.168 & 0.1500 & 8434 & $ 1226 \pm 26 $ & $ 1.12 \pm 0.02 $ & B \\
& 0.168 & 0.210 & 0.1880 & 7383 & $ 1478 \pm 32 $ & $ 1.09 \pm 0.02 $ & B \\
& 0.210 & 0.262 & 0.2351 & 21261 & $ 1720 \pm 31 $ & $ 1.19 \pm 0.02 $ & A \\
& 0.262 & 0.328 & 0.2932 & 20304 & $ 2068 \pm 41 $ & $ 1.09 \pm 0.02 $ & A \\
& 0.328 & 0.410 & 0.3665 & 18032 & $ 2440 \pm 51 $ & $ 1.03 \pm 0.02 $ & A \\
& 0.410 & 0.512 & 0.4577 & 15535 & $ 2862 \pm 62 $ & $ 1.00 \pm 0.02 $ & A \\
& 0.512 & 0.640 & 0.5717 & 13234 & $ 3309 \pm 35 $ & $ 0.98 \pm 0.01 $ & A \\
& 0.640 & 0.800 & 0.7140 & 10612 & $ 3629 \pm 46 $ & $ 0.96 \pm 0.01 $ & A \\
& 0.800 & 1.000 & 0.8913 & 8608 & $ 4101 \pm 56 $ & $ 0.96 \pm 0.01 $ & A \\
& 1.000 & 1.250 & 1.1157 & 6688 & $ 4477 \pm 69 $ & $ 0.96 \pm 0.01 $ & A \\
& 1.250 & 1.560 & 1.3932 & 4998 & $ 4599 \pm 76 $ & $ 0.94 \pm 0.01 $ & A \\
& 1.560 & 1.950 & 1.7374 & 3797 & $ 4961 \pm 90 $ & $ 0.97 \pm 0.01 $ & A \\
& 1.950 & 2.440 & 2.1728 & 2868 & $ 5141 \pm 105 $ & $ 0.95 \pm 0.01 $ & A \\
& 2.440 & 3.050 & 2.7193 & 1973 & $ 5031 \pm 121 $ & $ 0.96 \pm 0.01 $ & A \\
& 3.050 & 3.820 & 3.3995 & 1500 & $ 5398 \pm 148 $ & $ 0.98 \pm 0.01 $ & A \\
& 3.820 & 4.770 & 4.2397 & 1043 & $ 5263 \pm 175 $ & $ 0.98 \pm 0.01 $ & A \\
& 4.770 & 5.960 & 5.2943 & 627 & $ 4298 \pm 184 $ & $ 0.96 \pm 0.01 $ & A \\
& 5.960 & 7.450 & 6.6425 & 463 & $ 4670 \pm 238 $ & $ 1.00 \pm 0.02 $ & A \\
& 7.450 & 9.310 & 8.3012 & 305 & $ 4243 \pm 259 $ & $ 0.99 \pm 0.02 $ & A \\
& 9.310 & 11.600 & 10.3228 & 210 & $ 4150 \pm 286 $ & - & A \\
& 11.600 & 14.600 & 13.0154 & 138 & $ 3716 \pm 316 $ & - & A \\
& 14.600 & 18.200 & 16.1551 & 83 & $ 3197 \pm 351 $ & - & A \\
& 18.200 & 22.700 & 20.4050 & 70 & $ 3867 \pm 462 $ & - & A \\
& 22.700 & 28.400 & 24.7647 & 37 & $ 2618 \pm 430 $ & - & A \\
& 28.400 & 56.800 & 35.5961 & 50 & $ 1759 \pm 249 $ & - & A \\
& 56.800 & 113.700 & 75.5865 & 13 & $ 1500^{+543}_{-410} $ & - & A \\
\hline
88 & 0.262 & 0.328 & 0.2929 & 5937 & $ 2413 \pm 52 $ & $ 1.15 \pm 0.02 $ & B \\
& 0.328 & 0.410 & 0.3666 & 5329 & $ 3038 \pm 68 $ & $ 1.14 \pm 0.02 $ & B \\
& 0.410 & 0.512 & 0.4578 & 4802 & $ 3585 \pm 85 $ & $ 1.07 \pm 0.02 $ & B \\
& 0.512 & 0.640 & 0.5733 & 14344 & $ 3992 \pm 81 $ & $ 1.08 \pm 0.02 $ & A \\
& 0.640 & 0.800 & 0.7149 & 12701 & $ 4580 \pm 99 $ & $ 1.01 \pm 0.02 $ & A \\
& 0.800 & 1.000 & 0.8933 & 10527 & $ 5158 \pm 69 $ & $ 0.98 \pm 0.01 $ & A \\
& 1.000 & 1.250 & 1.1148 & 8722 & $ 5791 \pm 83 $ & $ 0.96 \pm 0.01 $ & A \\
& 1.250 & 1.560 & 1.3943 & 6864 & $ 6362 \pm 98 $ & $ 0.95 \pm 0.01 $ & A \\
& 1.560 & 1.950 & 1.7392 & 5309 & $ 6653 \pm 116 $ & $ 0.93 \pm 0.01 $ & A \\
& 1.950 & 2.440 & 2.1731 & 3947 & $ 6937 \pm 125 $ & $ 0.94 \pm 0.01 $ & A \\
& 2.440 & 3.050 & 2.7171 & 2947 & $ 7211 \pm 154 $ & $ 0.93 \pm 0.01 $ & A \\
& 3.050 & 3.820 & 3.3946 & 2131 & $ 7448 \pm 190 $ & $ 0.96 \pm 0.01 $ & A \\
& 3.820 & 4.770 & 4.2569 & 1537 & $ 7604 \pm 220 $ & $ 0.95 \pm 0.01 $ & A \\
& 4.770 & 5.960 & 5.3082 & 1061 & $ 7474 \pm 250 $ & $ 0.98 \pm 0.01 $ & A \\
& 5.960 & 7.450 & 6.5933 & 658 & $ 6216 \pm 261 $ & $ 0.95 \pm 0.01 $ & A \\
& 7.450 & 9.310 & 8.2753 & 471 & $ 6297 \pm 344 $ & $ 0.95 \pm 0.03 $ & A \\
& 9.310 & 11.600 & 10.3467 & 317 & $ 6218 \pm 367 $ & $ 0.99 \pm 0.02 $ & A \\
& 11.600 & 14.600 & 12.9277 & 227 & $ 6010 \pm 399 $ & - & A \\
& 14.600 & 18.200 & 16.2550 & 139 & $ 5437 \pm 461 $ & - & A \\
& 18.200 & 22.700 & 20.1788 & 81 & $ 4352 \pm 484 $ & - & A \\
& 22.700 & 28.400 & 25.7641 & 67 & $ 5235 \pm 640 $ & - & A \\
& 28.400 & 56.800 & 36.8540 & 86 & $ 3300 \pm 356 $ & - & A \\
& 56.800 & 113.700 & 78.1978 & 15 & $ 1884^{+624}_{-480} $ & - & A \\
\hline
\end{tabular}
\end{table*}

\end{appendix}

%%%%%%%%%%%%%%%%%%%%%%%%%%%%%%%%%%%%%%%%%%%%%%%%%%%%%%%%%%%%%%%%%%%
\bibliographystyle{pasa-mnras}
\bibliography{1r_lamboo_notes}
\setlength{\labelwidth}{0pt}

%%%%%%%%%%%%%%%%%%%%%%%%%%%%%%%%%%%%%%%%%%%%%%%%%%%%%%%%%%%%%%%%%%%
\label{lastpage}
\end{document}